\documentclass[10pt, conference, letterpaper]{IEEEtran}

\usepackage[pass]{geometry}

\usepackage{algorithm}
\usepackage{amsfonts}
\usepackage{amsthm}
\usepackage{amsmath}
\usepackage{array}
\usepackage{bbm}
\usepackage{cite}
\usepackage{dsfont}
\usepackage{epsfig}
\usepackage{float}
\usepackage[T1]{fontenc}
\usepackage{graphicx}
\usepackage{indentfirst}
\usepackage{multicol}

\usepackage{algorithmicx}
\usepackage{algpseudocode}
\newcommand\Algphase[1]{%
\vspace*{-.7\baselineskip}\Statex\hspace*{\dimexpr-\algorithmicindent-2pt\relax}\rule{\columnwidth}{0.4pt}%
\Statex\hspace*{-\algorithmicindent}\textbf{#1}%
\vspace*{-.7\baselineskip}\Statex\hspace*{\dimexpr-\algorithmicindent-2pt\relax}\rule{\columnwidth}{0.4pt}%
}

\usepackage{array}
\newcolumntype{L}[1]{>{\raggedright\let\newline\\\arraybackslash\hspace{0pt}}m{#1}}
\newcolumntype{C}[1]{>{\centering\let\newline\\\arraybackslash\hspace{0pt}}m{#1}}
\newcolumntype{R}[1]{>{\raggedleft\let\newline\\\arraybackslash\hspace{0pt}}m{#1}}

\usepackage{subfigure}
\usepackage{url}
\usepackage{multirow}
\usepackage{color}
\usepackage{blkarray} 
\usepackage{tikz}
\usetikzlibrary{calc}
\usepackage[export]{adjustbox} 
\usepackage{setspace}


\let\mathsetfont\mathcal

\newcommand\setD{\mathsetfont D}

\newcommand\setN{\mathsetfont N}

\newfont{\mycrnotice}{ptmr8t at 7pt}
\newfont{\myconfname}{ptmri8t at 7pt}

\IEEEoverridecommandlockouts

\begin{document}

\def\sharedaffiliation{%
\end{tabular}
\begin{tabular}{c}}



\title{DORE: An Experimental Framework to Enable Outband D2D Relay in Cellular Networks}

\author{\IEEEauthorblockN{Arash Asadi}
\IEEEauthorblockA{Technische Universit\"at Darmstadt\\
aasadi@seemoo.tu-darmstadt.de}
\and
\IEEEauthorblockN{Vincenzo Mancuso\thanks{This manuscript is an extended version of the work accepted for
presentation at IEEE INFOCOM 2016~\cite{asadi2016INFOCOM}.}}
\IEEEauthorblockA{IMDEA Networks Institute\\
vincenzo.mancuso@imdea.org}
\and
\IEEEauthorblockN{Rohit Gupta}
\IEEEauthorblockA{Actility \\
rohit.gupta@actility.com }
}



%

\maketitle

\begin{abstract}

Device-to-Device communications represent a paradigm shift in cellular networks. In particular, analytical results on D2D performance for offloading and relay are very promising, but no experimental evidence validates these results to date. This paper is the first to provide an experimental analysis of outband D2D relay schemes.  Moreover, we design DORE, a complete framework 
for handling channel opportunities offered by outband D2D relay nodes. DORE consists of resource allocation optimization tools and protocols suitable to integrate
QoS-aware opportunistic D2D communications within the architecture of 3GPP {\it Proximity-based Services}. We implement DORE using an SDR framework to profile cellular network dynamics in the presence of opportunistic outband D2D communication schemes. 
Our experiments reveal that outband D2D communications are suitable for relaying in a large variety of delay-sensitive cellular applications, and that DORE enables notable gains even with a few active D2D relay nodes.

\end{abstract}

\section{Introduction}
\label{s:intro}

Device-to-Device (D2D) communications gained rapid traction in academia and industry in recent years.
The popularity of D2D is due to its potential for solving a large spectrum of pressing issues  in today's cellular networks, e.g., insufficient capacity and lack of solutions for public safety applications. Indeed, 
a plethora of different studies have sprouted from the D2D research niche~\cite{kim2015TON, karvounas2014ComMag, bao2013Infocom, golrezaei2012Globecom,Liu2015mobicom}, which all agree on the crucial role of D2D in upcoming wireless systems.
Analytical and simulation-based results of these studies demonstrate outstanding  gains for applications like opportunistic relay, cellular offloading, and cell coverage extension, especially under opportunistic channel utilization~\cite{jiang2016JSAC,hourani2016TMC, ji2016JSAC}. 
However, D2D schemes are tightly integrated with the cellular infrastructure, which is a rare commodity in academia, hence the lack of experimental evaluations.

3GPP  is actively studying  the feasibility and the architecture of D2D communications to finalize the standardization process for both {\it inband} and {\it outband} D2D modes, in which inband D2D uses the cellular spectrum, while outband D2D uses unlicensed spectrum~\cite{asadi2014MAMA}. 
More in general, 
the state-of-the-art clearly shows that inband D2D is a well-explored topic~\cite{mach2015Survey}. However, its standardization is progressing slowly due to the significant modifications required for accommodating D2D users in the cellular spectrum. 
In contrast, outband D2D communications do not require a significant change in the resource management of the cellular spectrum, which explains why they are now receiving more attention~\cite{andreev2014ComMag}.
The pivotal technological challenge to implement outband D2D schemes 
consists in the User Equipment (UE)-relay feature, which undergoes an extensive investigation in the 3GPP's study on architectural enhancements to support {\it Proximity-based Services} (ProSe)~\cite{3GPP23.703,lin2014ComMag}. With the above, it will be possible to tackle D2D use-cases such as content sharing, offloading, and public safety using the unlicensed spectrum to assist cellular operation.
%

This is the first work to build an SDR-based experimental testbed for outband D2D communications. In particular, we leverage the experimental setup for rigorous tests investigating the performance and practicality of channel-opportunistic outband D2D communications under tight QoS constraints.
 The following list summarizes our contributions:
$(i)$ We design D2D Opportunistic Relay with QoS Enforcement (DORE), a channel-opportunistic framework for enhancing network capacity under QoS constraints;
$(ii)$ We formulate DORE as a QoS-aware throughput maximization problem to perform relay and mode selection for ProSe-enabled UEs. In this paper, a UE is either in outband D2D mode or legacy cellular mode. Inband D2D is out of our scope;
$(iii)$ We design a greedy algorithm for implementing DORE, based on the aforementioned problem formulation for time-stringent operations;
$(iv)$ We design a protocol to integrate DORE into the 3GPP ProSe architecture; 
$(vi)$ We point out a few shortcomings of ProSe and propose new amendments;
$(v)$ Using an SDR platform and commercial-off-the-shelf Android devices, we implement the first experimental testbed for DORE and, more in general, for outband D2D and opportunistic outband D2D solutions;
$(vi)$ We evaluate outband D2D and DORE  with real-time video streaming and non-delay-tolerant flows.


Our results indicate that outband D2D is indeed a feasible scheme that is suitable for a large variety of cellular applications. We also show that not only channel quality-based opportunities exist in abundance in a cellular network, but also it is feasible to design simple schemes to leverage such opportunities efficiently. 
\begin{figure*} [t!]
\centering
		\includegraphics[scale=0.5, angle=0]{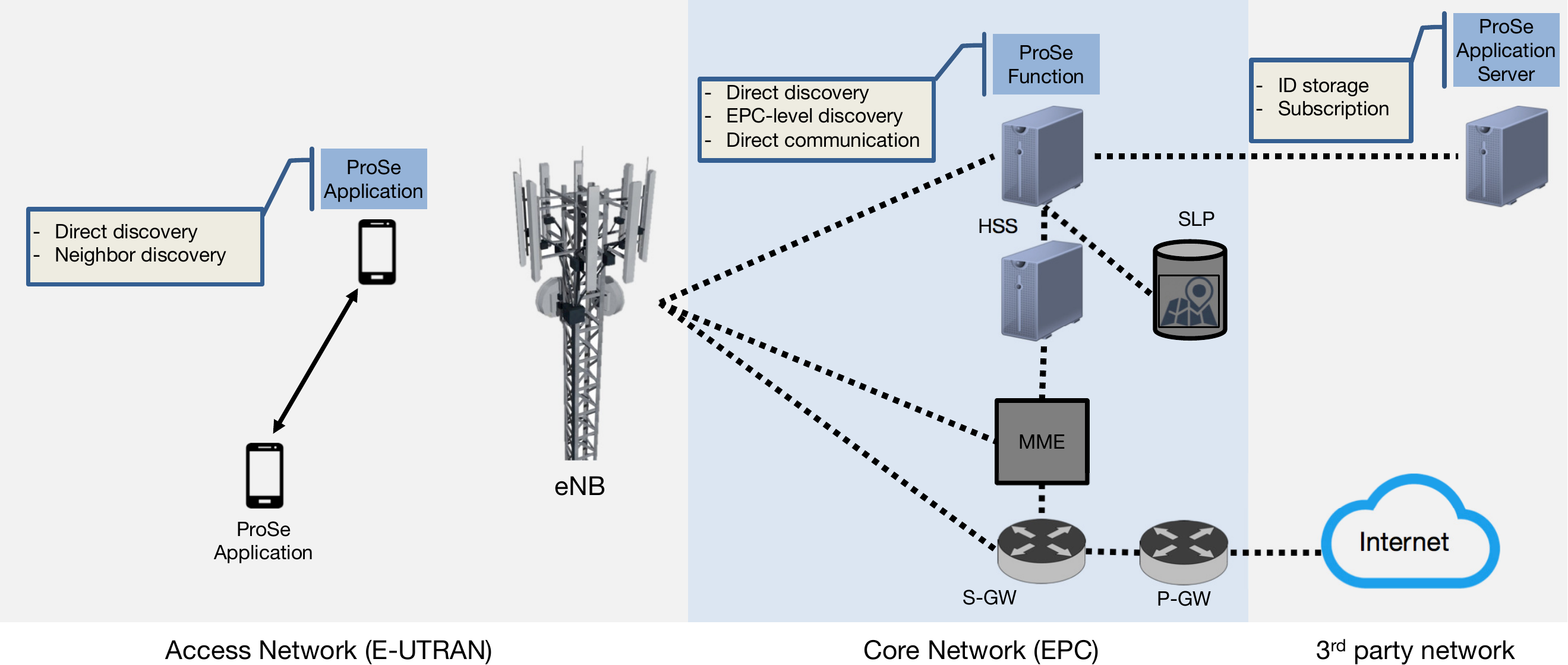} 
		\vspace{-2mm}
		\caption{3GPP's envisioned ProSe architecture. ProSe Application which runs at the user side is located at the Access network. The ProSe Function is in the EPC, in the core of the cellular network, and  acts as an anchor between ProSe Application and ProSe Application Server. The ProSe Application Server can be in the core network as well as  outside the operator infrastructure, in a third party network.}
		\vspace{-4mm}
		\label{fig:proseArch}
\end{figure*}

The organization of this article is as follows. Section~\ref{s:proto} provides a short background on 3GPP ProSe architecture and introduces the problem formulation, our proposed greedy algorithm and 3GPP ProSe-compliant protocol for DORE. In Section~\ref{s:setup}, we elaborate on details of our SDR testbed and its setup. The results of the experiments are reported in Section~\ref{s:eval}. We discuss the lessons learned in Section~\ref{s:disc}. The related work is commented in Section~\ref{s:related}. Finally, Section~\ref{s:conclusions} concludes this article.



\section{DORE}
\label{s:proto} 

Despite the unavailability of standardization for D2D communications, the possible D2D architecture designs could be deduced from the ongoing feasibility studies and technical reports that deal with D2D (see in particular~\cite{3GPP23.703, 3GPP23.303, 3GPP36.843}). 
In this section, we design DORE, a framework for {\it signal-quality based} opportunistic outband D2D relays, as the key enabler technology for improving wireless users' performance. DORE builds on an optimization problem that is formulated for relay/mode selection with delay constraints. Given the NP-Hardness of the optimization problem, we design a greedy algorithm for DORE. Moreover, 
to evaluate the performance of DORE, we develop a practical protocol that leverages and amends the existing 3GPP ProSe framework.
In this article, we assume that UEs can communicate using Legacy Cellular mode or Outband D2D mode. 

\subsection{Background on 3GPP ProSe Architecture}
There are three main elements in 3GPP for supporting  D2D communications, namely, ProSe Application, ProSe Application Server, and ProSe Function.  Fig.~\ref{fig:proseArch} graphically depicts the location and interworking of each element in 3GPP's architecture. In addition, other existing network entities such as Mobility Management Entity (MME) and Home Subscriber Server (HSS) are amended to support communication with these new ProSe-specific elements. MME is responsible for initiating paging, authentication and tracking the location of the mobile device. HSS is a database that contains the subscriber's information. 
Here, we briefly describe the main ProSe elements and refer the interested reader to~\cite{3GPP23.303} for more details.

\begin{itemize}
\item {\bf ProSe Application}. The ProSe Application runs on the ProSe-enabled UEs and supports control and data communication between ProSe-enabled UEs and the ProSe Function. Moreover, direct discovery procedure is handled by ProSe Applications. For example, a given ProSe Application can send beacons to discover new nodes or it can report its location information to the Evolved Packet Core (EPC) to be discovered through network assistance.

\item {\bf ProSe Function}. The ProSe Function is a logical function located inside EPC to handle network related procedures of ProSe. In the current specification~\cite{3GPP23.303}, ProSe Function may play different roles depending on the use-case. ProSe Function mainly assists in direct discovery, EPC-level discovery, and direct communication. The location information used for EPC-level discovery is obtained from the SUPL Location Platform (SLP), to which UEs periodically inform their location (see Section 5.5.6 in~\cite{3GPP23.303} for more information). ProSe function also acts as a reference point between UEs and Application Server.

\item {\bf ProSe Application Server} This entity is located outside the EPC and it provides the necessary functionalities for the ProSe specific operations. It also supports application layer functionalities such as the storage of ProSe User and ProSe Function IDs, and mapping between them.

\end{itemize}

The key procedures in D2D communications are Discovery and Direct Communication. 

{\bf Discovery} is the procedure that allows a UE to discover another UE before the initiation of communication. According to 3GPP, there are two types of discovery, namely, direct discovery and EPC-level discovery. In direct discovery, UE searches for other devices in proximity independent of the network. Here, the UEs use periodical beacon transmission/reception to communicate with other UEs. This approach is particularly beneficial in out-of-coverage scenarios where network assistance is no longer available. In EPC-level discovery, as the name suggests, the EPC keeps track of the UE's location and triggers the device discovery process upon detection of another UE in proximity. 
This approach reduces the energy consumption at the UE since the transmission of the discovery beacons is not required.

Upon the completion of the discovery procedure between the UEs, {\bf Direct Communication} is the procedure that allows inter-UE communications, which can be network-authorized or network-independent. Both methods support one-to-one, one-to-many, and relay use-cases. In addition, communication can occur inband (over the licensed spectrum) or outband (over unlicensed spectrum).

\subsection{System Model} 

In DORE, UEs establish outband D2D links using WiFi in unlicensed bands and exploit the channel diversity of cellular networks opportunistically. To this aim, the eNB evaluates the reported Channel Quality Indicators (CQIs) of the D2D UEs and selects the UE with the highest CQI as the relay. Therefore, at each scheduling epoch, a relay node receives its traffic plus the traffic for other UEs (i.e., D2D receivers) with lower channel qualities. Also, at each scheduling interval, the roles (i.e., relay or receiver) can be switched according to the reported CQI values. Nevertheless, if the use of outband D2D link results in the violation of QoS constraints, the UEs fall back to cellular mode. Note that a relay node can serve multiple receivers simultaneously. 

\begin{figure} [t!]
\centering
		\includegraphics[scale=0.55, angle=0]{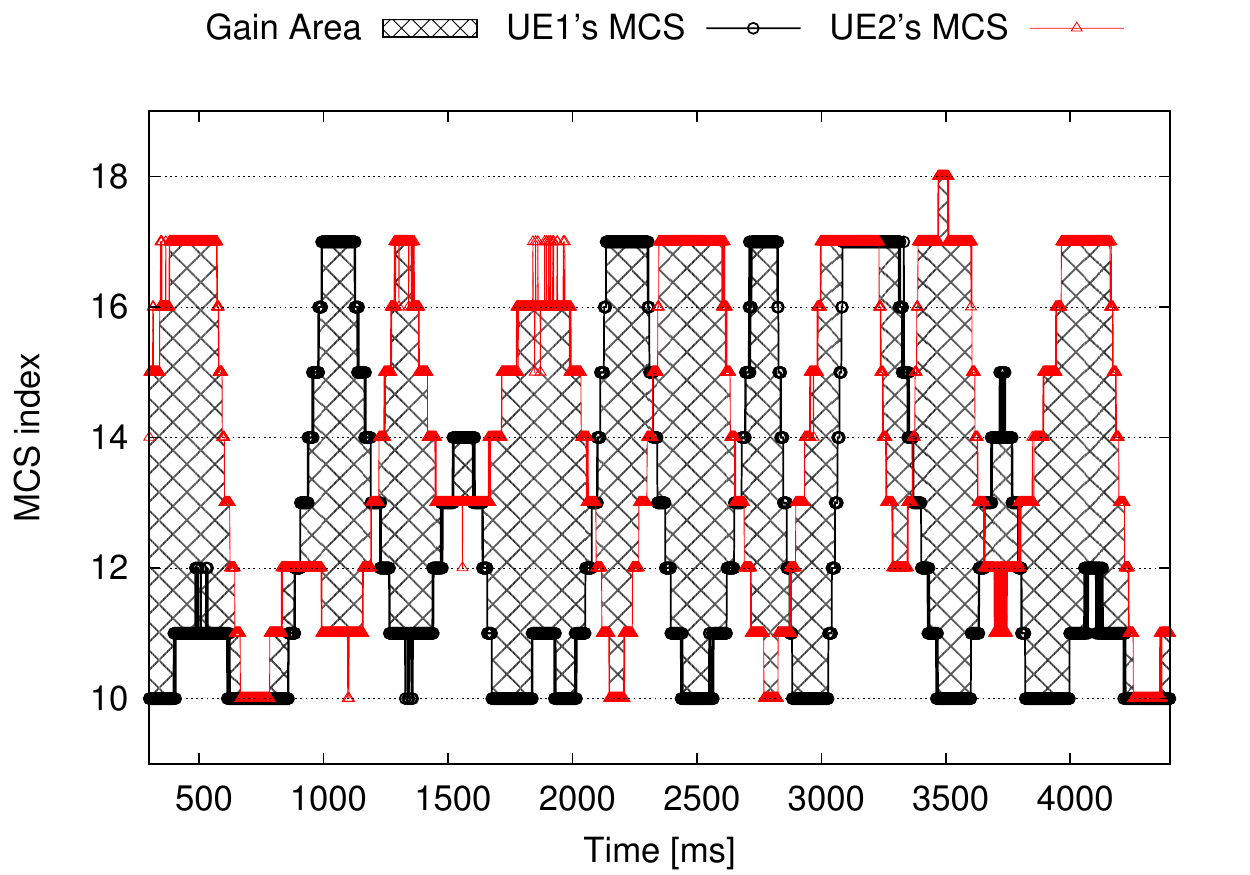} 
		\caption{A snapshot of MCS indices for two UEs, and their {\it Gain Area}. Using an opportunistic approach, users can always communicate with the eNB with the highest MCS. This is done by simply allowing the UE with higher MCS to relay for the other UE.}
		\label{fig:2usrGain}
\end{figure}

The described opportunistic outband D2D relay scheme is compatible with any scheduling mechanism employed at the eNB. In particular, in DORE, the eNB scheduler allocates the resource blocks to the UEs connected via D2D in the same manner as 
for legacy cellular UEs, except that: $(i)$ Modulation and Coding Scheme (MCS) of the D2D receiver is adapted to the MCS of the UE acting as relay, and $(ii)$ the relay receives the traffic of the D2D receiver over cellular interface and forwards it over the D2D link. 
The opportunism in DORE is better illustrated in Fig.~\ref{fig:2usrGain}, where we show a snapshot of the best MCSs usable by UE1 and UE2 over an interval of $\sim 4s$. 
We can observe that the MCS of the UEs is often different. The opportunistic approach allows for communicating with the highest MCS available to the two UEs. For example when UE1 has a low MCS and UE2 has a high MCS, DORE will relay UE1's traffic through UE2, whose MCS is the highest. In Fig.~\ref{fig:2usrGain}, we demonstrate this potential gain in the hatched area enclosed between the two curves (i.e., the {\it Gain Area} indicated in the figure). Clearly, opportunistic outband D2D is beneficial in scenarios with low channel correlation among users.

To get a better grasp of the potential gain in a real network, we measure the channel quality of $15$ smartphones (all subscribed to the same operator) connected to the same eNB. The measurement is done in our premises in Madrid in an office area of approximately $500$ $m^2$ for the period of $12$ hours. Fig.~\ref{fig:clusteringGain} shows the results in terms of average transport block size in bits. In LTE terminology, a transport block is the data send from MAC layer to physical layer. The duration of a transport block is $1$ ms, and the MCS dictates the number of bits to be transmitted within that block. In this figure, we illustrate the average transport block size when all UEs form one group, when all UEs are divided into five groups, and finally when all UEs operate individually (i.e., $x=15$). The error-bars are the $25^{th}$ and the $75^{th}$ percentiles. The results demonstrate that once UEs are divided into five groups (i.e., three UEs per group), the average transport block size improves by $18.4\%$. We deem this result very motivating because opportunistic D2D relay can be effective even when all UEs observe very high channel quality (64QAM). Moreover, the performance improves with as few as five collaborating UEs.  

\begin{figure} [t!]
\centering
		\includegraphics[scale=0.55, angle=0]{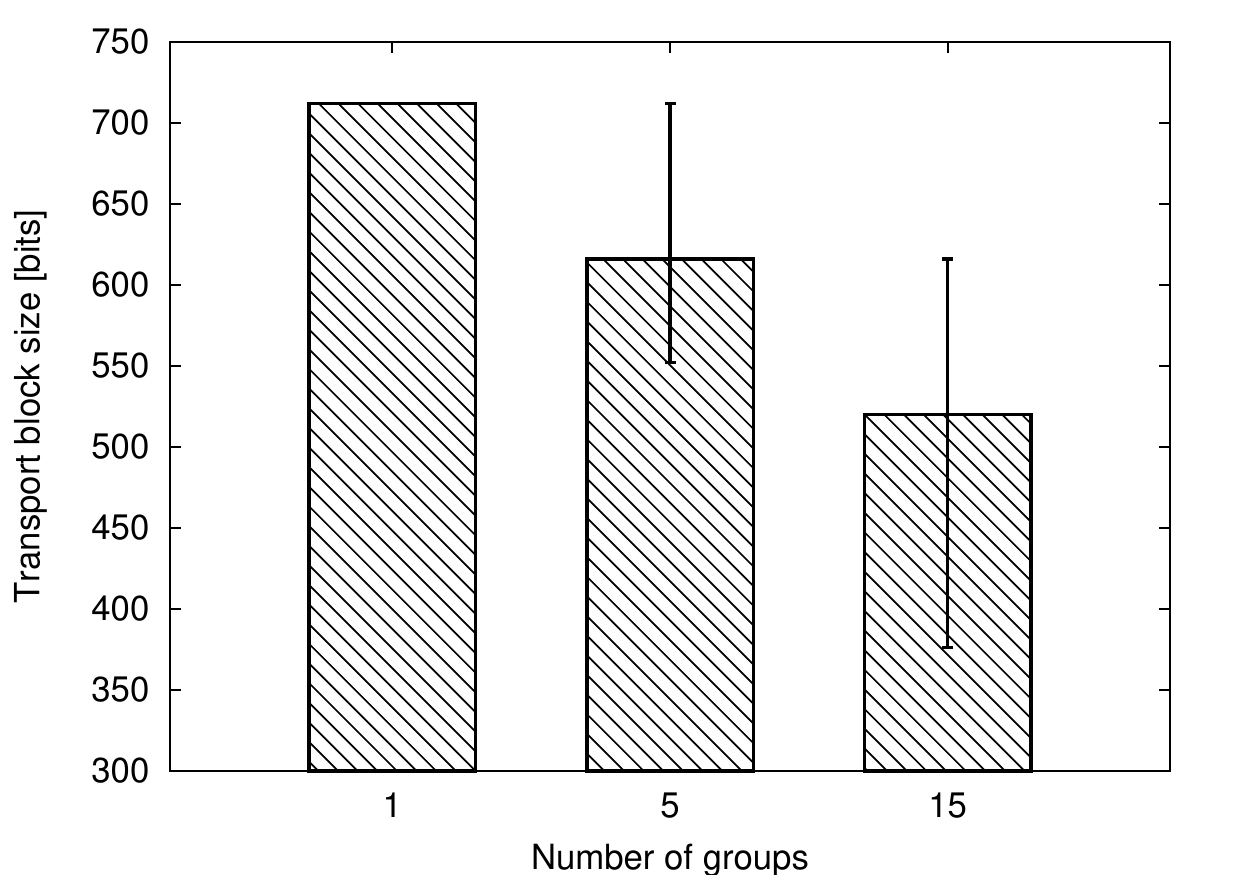} 
		\caption{Impact of opportunistic D2D relay based on real-time traces of channel qualities of 15 cellular users. The figure shows the average number of bits that can be packed into one transport block.} 
		\label{fig:clusteringGain}
\end{figure}

\subsection{Problem Formulation} 

One of the main concerns about outband D2D is maintaining the QoS requirements of the flows. Hence, we formulate an optimization problem for 
DORE to maximize system throughput by performing relay/mode selection under delay constraints.

The ProSe Server is aware of the throughput and the location of the users since the UEs report their location and link quality as specified by 3GPP ProSe technical specification (see Sections 5.4.4 and 5.5.6 in~\cite{3GPP23.303} for more information).

The throughput (in bits/sec) of a user $i$ who receives its data directly from the eNB can be expressed as 
$T_{ lte }^{ ( i ) }= r_{ lte } ^ { ( i ) } b_{ lte } ^{ ( i ) }$, 
where $r_{ lte }^{ ( i ) }$ is the symbol rate (bits/sym) and 
$b_{ lte }^{ ( i ) }$ is the allocated bandwidth in terms of number of symbols per second (sym/sec). 
Alternatively, the throughput of a user~$i$ that receives its traffic from a relay $j$ is expressed as 
$T_{ d2d }^{ ( ij ) }= \min ( r_{ lte }^{ ( j ) }  b_{ lte }^{ ( i ) }, r_{ wifi }^{ ( ij ) } b_{ wifi }^{ ( i ) } ) $,
where $r_{ lte }^{ ( j ) }$ and $r_{ wifi }^{ ( ij ) }$ are the achievable rates of relay over LTE and the achievable WiFi rate between the relay $ j $ and the receiver $i$, respectively. $b_{ wifi }^{ ( i ) }$ is the available bandwidth to user $ i $ over WiFi. $b_{ lte }^{ ( i ) }$ is the available bandwidth to user $ i $ but with the rate matching the channel quality of user $j$, $r_{ lte }^{ ( j ) }$. Naturally, $T_{ d2d }^{ ( ij ) }$ is the minimum between the throughput of the two links (i.e.,  LTE and WiFi). This is similar to classic routing scenarios where the capacity of a segmented path is capped by the link with the lowest capacity. 
Given these definitions, 
and using $\mathcal{N}$ to denote the set composed by an eNB and $N$ users,
we can formulate 
DORE  as a throughput maximization problem with delay constraint $d_{ th }^{ ( i ) }$ for $N$ users as follows:

\begin{align}
\begin{cases}
\max	 	& \sum_{ i\in\setN\setminus \{0\} } \sum_{ j\in\setN\setminus \{i\} }  ( \alpha_{ ij } T_{ lte }^{ ( i ) } +\alpha_{ ji } T_{ d2d }^{ ( ij ) } )  \label{eq:max}\\
\text{s.t.} 	& \alpha_{ ij }\in\{0,1\} \quad \forall i,j\in\{0,\cdots,N\} \\
		& \alpha_{ ii } = 0 \quad i\neq 0 \\
		& \alpha_{00} = 1 \\
		&  \alpha_{ i0 } = 0  \qquad \forall i \in\{1,\cdots,N\} \\
		& \sum_{ i\in\setN } \alpha_{ ij } = 1 \qquad \forall j \in\{1,\cdots,N\}  \\
				& \sum_{ j\in\setN } \alpha_{ ij } - \left( N\!-\!1 \!+\! 2\delta_{ 0i } \right) \alpha_{ 0i } \leq 0  \quad \forall i \in\{0,\cdots,N\}  \\ 		
		& d^{ ( ij ) } \leq d^{ ( i ) }_{ th }
\end{cases}
\end{align}

\noindent where $\alpha_{ ij }$ is a binary decision variable that determines whether user $i$ transmits to user $j$ or not. For notational convenience, we mark the eNB as user $0$. For notational and formulation convenience, we set $\alpha_{ 00 } = 1$. This allows to differentiate transmitters---either eNB or relays---from final destinations users (UEs) in the sixth constraint above, and it does not project any physical meaning. 


More in detail, the first constraint of~\eqref{eq:max} forces the decision variable to be binary. The second constraint avoids a user transmits to itself. The third constraint ensures that the eNB is active. 
The fourth constraint excludes transmissions to the eNB. The fifth constraint enforces the users to receive data either from the eNB or a relay. The sixth constraint limits the number of receivers of a relay to up to $N-1$ and the transmissions from the eNB to $N$ (note that $\alpha_{00}$ is $1$, but it does not count as a transmission from the eNB). Here, we use $\delta_{ij}$ as an assisting parameter to formulate the fact that the eNB is allowed to transmit to all of the $N$ users and to limit the relays to transmit only to other UEs. To this aim, $\delta_{ij}$ is $1$ for $i=j$ and $0$ otherwise.
Finally, the last constraint keeps the total delay $d^ { ( ij ) }$ below a threshold $d_{th}^ { ( i ) }$. Here, $d^ { ( ij ) }$ is measured as the time taken for a packet to reach from the eNB to a user $ i $ via relay $ j $ ($j=0$ when there is no relay involved). 

\subsection{Complexity} 

Problem~\eqref{eq:max} is a Knapsack problem 
with multiple dimensions, because of the presence of multiple delay constraints, each of which can be seen as a dimension of the knapsack, in addition to the fact that we cannot accommodate more than $N$ nodes in the group of relay nodes. This class of problems 
is proven to be NP-Hard \cite{caprara2014SIAM}
 and has no Efficient Polynomial-Time Approximation Scheme (EPTAS) but for some polynomial time algorithms that provide performance guarantees, in terms of a bound on the distance from the optimal, for {\it sparse} cases only~\cite{CG14}.
Although the complexity of the brute force algorithm is   O$\left(2 ^ {N^{2 -1} }\right)$, the set of feasible solutions is significantly smaller than the original set. The following shows the set of decision variables assuming that $m$ users directly communicate with the eNB (i.e., relay/cellular UE):

 \definecolor{orange}{rgb}{0.05,0.55,1}
\newcommand\hlight[1]{\tikz[overlay, remember picture,baseline=-\the\dimexpr\fontdimen22\textfont2\relax]\node[rectangle,fill=orange!100,rounded corners,fill opacity = 0.2,fill,thick,text opacity =1] {$#1$};}

\newcommand{\tikzmark}[1]{\tikz[overlay,remember picture] \node (#1) {};}
\newcommand{\DrawBox}[1][]{%
    \tikz[overlay,remember picture]{
    \draw[red,#1]
      ($(left)+(-0.4em,0.7em)$) rectangle
      ($(right)+(0.2em,-0.3em)$);}
}

\begin{equation}
\resizebox{0.98\columnwidth}{!}	{$
 \begin{blockarray}{cccccccc}
 			&					&\BAmulticolumn{3}{c}{relay/cellular~UE}	&\BAmulticolumn{3}{c}{receiver}	\\
     			& eNB 				& u_{ 1 }			& \dots  	&u_{ m } 			&u_{ m+1 } 		& \dots 	&u_{ N }\\
    \begin{block}{c[c|ccc|ccc@{\hspace*{-1pt}}]}
    eNB 		& \alpha_{00}=1  			& \alpha_{01}=1 	& \dots  	& \alpha_{0m}=1 	& \alpha_{0m+1}=0  &\dots	&\alpha_{0N}=0 \\
    \cline{2-8}
    u_{ 1 }		& \alpha_{10}=0 			& \tikzmark{left} \alpha_{11}=0 	& \dots  	& \alpha_{1m}=0	& \hlight{\alpha_{1m+1}=?}  &\hlight{\dots} 	&\hlight{\alpha_{1N}=?} \\
    \vdots		&\vdots 				& \vdots 			& \vdots 	& \vdots			& \hlight{\vdots}			& \hlight{\vdots} 	& \hlight{\vdots}\\
    u_{ m }		& \alpha_{m0}= 0			& \vdots			& \vdots  	&  \vdots		 	& \hlight{\alpha_{mm+1}=?}  &\hlight{\dots} 	&\hlight{\alpha_{mN}=?} \\
    \cline{6-8}
    u_{ m+1 }	& \alpha_{m+1 \; 0}=0 		&  \vdots			& \vdots  	&  \vdots		 	& \alpha_{m+1m+1}=0  &\dots 	&\alpha_{m+1N}=0 \\
    \vdots		&\vdots 				& \vdots 			& \vdots 	& \vdots			& \vdots			& \vdots 	& \vdots\\
    u_{ N }		& \alpha_{N 0} = 0			& \alpha_{N1}=0	& \dots  	& \alpha_{Nm}=0\tikzmark{right} 	& \alpha_{Nm+1}=0  &\dots 	&\alpha_{NN}=0 \\
    \end{block}
      \DrawBox[thick]
  \end{blockarray}\nonumber
  $}
\end{equation}

\noindent For the sake of presentation, we sorted the matrix so that the relays and receivers appear in two distinct groups.  All elements in the first column, but $\alpha_{00}$, are zeros because users cannot relay to the eNB. In the first row, all elements are ones but the D2D receivers because they are not supposed to communicate with the eNB directly. Since each user can only receive from one source, columns $1$ to $m$ (in the red box) are all zeros but the first row elements. Finally, rows $m+1$ until $N$ are all zeros because D2D receivers can not transmit to other users. Now, we can see that, using the constraints defined in the problem, the complexity of the actual problem is reduced to a much smaller set of decision variables that is highlighted in light blue color in the top rightmost corner of the matrix. This reduces the complexity of the brute force to O$\left( \binom {N} {m} 2^{{(N-m)}^2}\right)$.
Nevertheless, the complexity is still high for real-time operation. Hence, we design a {\it greedy} algorithm based on the properties of the formulated problem. 


\subsection{A Greedy Algorithm for DORE}

The exact solution to Problem~\eqref{eq:max} is computationally expensive and does not allow for a rapid relay and mode selection. 

In addition, the fact that the multi-dimension knapsack problem has no EPTAS, but for sparse cases---and for our D2D case we cannot guarantee the {\it sparseness} of the problem because each node can potentially be a relay or ask to connect to a relay---makes it impossible to derive an algorithm to be at the same time efficient and optimal. Since our goal is to make the D2D relay approach implementable in real networks, we opt for a low complexity algorithm design and
propose a greedy algorithm (\texttt{Greedy}) in which we leverage the properties of our problem formulation to reduce the complexity of finding a solution. In line 1, as shown Algorithm~\ref{social}, \texttt{Greedy} obtains throughput and delay information of the D2D users from the eNB. The potential throughput gain $T_{ gain }^{ ( ij ) }$ in different configurations is computed in the nested {\it for loops} from lines 2 to 7.
In the second part (lines 8 to 16), \texttt{Greedy} starts the relay/mode assignment from the  $ij$ pair with the highest $T_{ gain }^{ ( ij ) }$ under delay constraint. With this implementation, we reduce the complexity to O($N^2$). 
Our greedy approach prioritizes the links with a higher potential D2D gain over those with lower gain. This opportunistic mode/relay selection results in a high D2D gain and prevents the users with good cellular links or poor D2D links from switching to D2D mode.
\texttt{Greedy} cannot be optimal because it runs in polynomial (quadratic) time. However, our experiments with low-to-medium sizes of D2D relay populations (no more than ten relay nodes) show that, on average, a greedy algorithm often achieves the optimal or at least very good results, a few percent below the optimal throughput. However, in some cases, the difference can be substantial, which was expected because of the multidimensional nature of the knapsack problem we have to address.

\begin{algorithm}[h]
\caption{ \texttt{Greedy}}
\scriptsize
\begin{algorithmic}[1]
\Require \\
		$ T_{ lte }^{ ( i ) }, T_{ d2d }^{ ( ij ) }, d^{ ( ij ) } \quad \forall i , j \in \{ 0 , \dots , N \} $.
\Ensure $\alpha_{ ij }$\\
initialize: $\alpha_{ 00 } = 1 $ , $\alpha_{ ij } = 0  \quad \forall i , j \in \{ 1, \dots , N \}$, $T_{ gain }^{ ( ij ) }=0$, $\setD=\emptyset$.
\For{$ i   \in \setN$ }
	\For{$ j   \in \setN\setminus{i}$ }
		\State{ $T_{ gain }^{ ( ij )} =T_{ d2d }^{ ( ij ) }-T_{ lte }^{ ( i ) }$}
	\EndFor
\EndFor

\For { k from 1 to N }
	\State{ find  $\arg\!\max_{(ij)}~T_{ gain }^{ ( ij ) } ,\quad i \in \setN \setminus \setD$ }
		\If {$d^{ ( ij ) }\leq d_{ th }^{ ( i ) } $}
			\State{ $ \alpha_{ ij }=1 \quad \& \quad \alpha_{ 0i }=1 $ }
			\State { $ \setD=\setD \cup \{ i \} $}
		\Else
			\State{ $ T_{ gain }^{ ( ij )} =0$ }	
		\EndIf	
\EndFor
\end{algorithmic}
\label{social}
\end{algorithm}

\subsection{DORE Procedures in a ProSe-Compliant Framework}
This subsection elaborates on the integration of DORE in 3GPP ProSe and our proposed amendments. With a few exceptions, our proposed protocol follows the ProSe specified procedure. For clarity, we elaborate on these exceptions in Section~\ref{ss:ProSeamend}. Fig.~\ref{fig:proto} illustrates the access network and main ProSe elements (i.e., ProSe Function and ProSe Application Server). The functionalities of these elements are designed to support a large spectrum of use-cases that makes ProSe very receptive to new protocols, including DORE. 
\begin{figure} [t!]
\centering
		\includegraphics[width=\columnwidth]{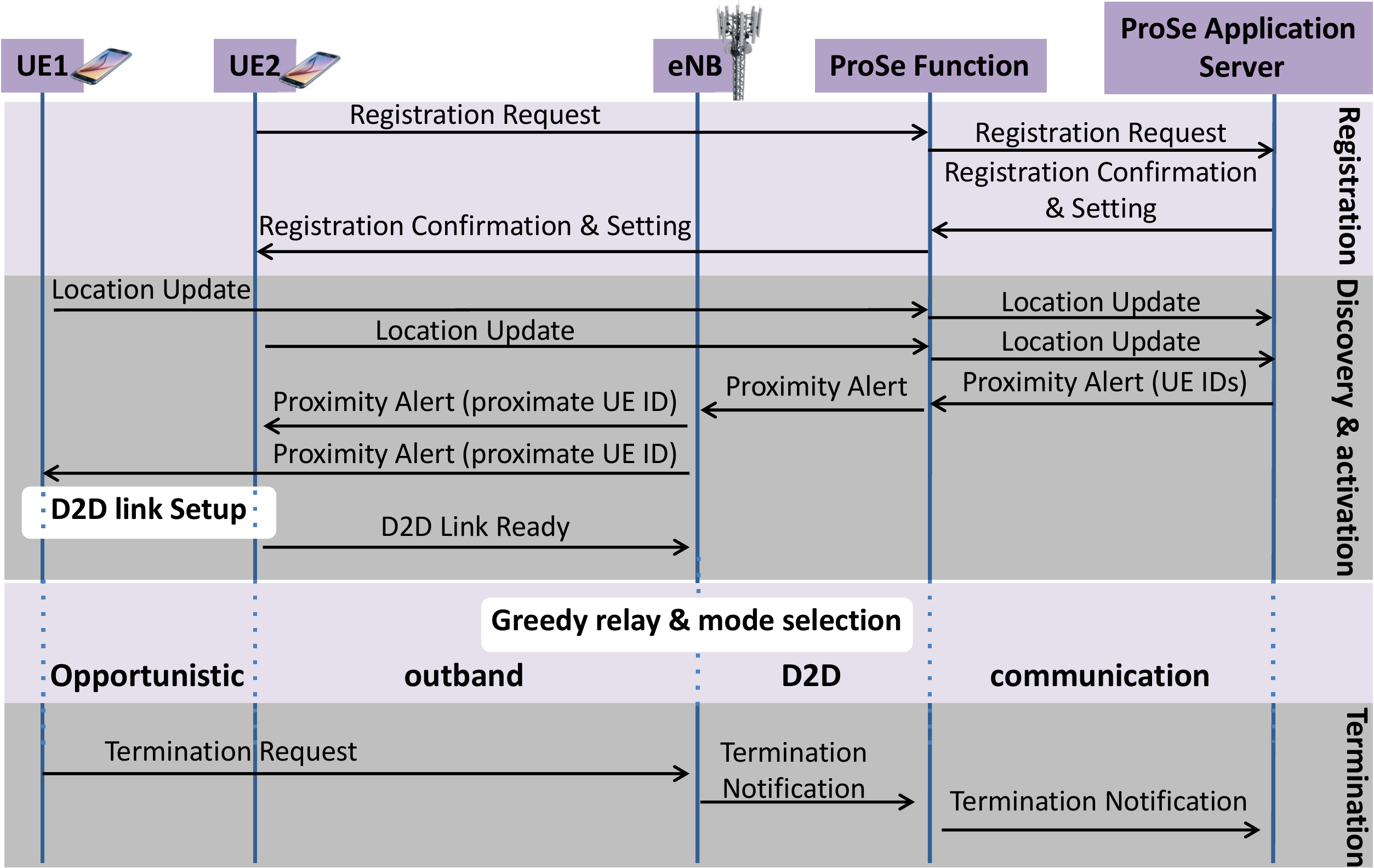} 
		\caption{Schematic protocol overview of DORE from registration phase to termination of the connection. In this figure, we assume that UE1 is already registered.}
		\label{fig:proto}
\end{figure}

\subsubsection{Registration}
UEs register for opportunistic outband D2D at ProSe Function by sending the {\it Registration Request} message, as shown in~Fig.~\ref{fig:proto}. This is necessary for the majority of D2D applications due to the operator-centric nature of D2D communications in cellular networks.  Note that in Fig.~\ref{fig:proto}, we assumed that UE1 is already registered.
ProSe Function responds to this request with a {\it Registration Confirmation and Settings} message. This message includes an application ID assigned to the UE for the requested service. The settings specify the periodicity of location updates and discovery beacon, and the discovery channel for the direct discovery method.

\subsubsection{Discovery}
Like any other D2D application, UEs can find other UEs in proximity using network-assisted discovery as illustrated in Fig.~\ref{fig:proto} or independently (i.e., direct discovery).

\indent {\bf EPC-level discovery.} In this mode, ProSe Function keeps track of the location of UEs that are registered for Opportunistic outband D2D service (based on the ProSe defined location reporting~\cite{3GPP23.703}). The registered UEs send the updated location information to ProSe Function at the intervals defined in the service setting received upon registration. Once two registered UEs are in proximity, ProSe function initiates the activation phase. \\
\indent {\bf Direct discovery.}  Prose Function informs the UEs on the WiFi channel to be used in discovery phase upon registration. In this mode, UEs use beacon transmission for active discovery or listening for a beacon on the discovery channel for passive discovery. 
This procedure resembles the discovery phase in WiFi Direct~\cite{WifiDirect2013}. The beacons are short frames that include UE's application ID and the type of application used.

\subsubsection{Activation}
Next, the system should establish a path between the eNB and D2D UEs. Since the relay-UE pairing dynamically changes in DORE, path reconfiguration from eNB to the destination UE should be quick and uninterrupted.  As a result, we propose to take the relay selection decision at the eNB instead of the ProSe Function and use D2D links that have been previously set up soon after their discovery. This is further elaborated in Section~\ref{ss:ProSeamend}.
The following details this procedure. 

\floatname{algorithm}{Activation} 
\renewcommand{\thealgorithm}{}  
\begin{algorithm}[H]
\caption{ }
\footnotesize 
\begin{algorithmic}[1]
\State ProSe Application Server sends {\bf Proximity Alert} message to the Prose Function. This message contains the application user ID of the D2D UEs in proximity.
\State ProSe Function sends {\bf Proximity Alert} message to the eNB and the UEs. The message to the eNB contains the UE cellular IDs while the message to the UE contains user application IDs to be used for D2D link activation.
\State UEs continue the link activation procedure per WiFi Direct standard.
\State Upon successful establishment of the connection, the UEs send {\bf D2D Link Ready} message to the eNB.
\end{algorithmic}
\label{activation}
\end{algorithm}

\subsubsection{Communication}
Once the eNB is notified on the D2D link activation, it starts to serve the D2D UEs based on our proposed \texttt{Greedy} algorithm for DORE as described below. 

\floatname{algorithm}{Communication} 
\renewcommand{\thealgorithm}{}  
\begin{algorithm}[H]
\caption{ }
\footnotesize 
\begin{algorithmic}[1]
\vspace{2mm}
\Algphase{Frame relay}
\State The \texttt{Greedy} algorithm performs D2D relay/mode selection at the eNB based on the delay and throughput feedback from the D2D UEs.
\State The eNB labels the frames of the D2D UEs so that each UE can differentiate if a received frame is local or it should be relayed. 
\State Upon reception of a relay frame, the relay UE processes the packet from the physical layer up to Packet Data Convergence Protocol (PDCP) layer. 
\State The relay UE encapsulates PDCP Service Data Unit (SDU) in a WiFi frame and forwards it over the D2D link. 
\State The D2D receiver decapsulates the relayed frame and processes the PDCP SDU through the rest of the LTE stack. 
\Algphase{Periodic updates}
\State UEs send regular CQI reports to the eNB for scheduling purposes. The periodicity is determined by the eNB.
\State UEs send the average achievable throughput and delay of the D2D link to the eNB for the Greedy algorithm. 
\end{algorithmic}
\label{activation}
\end{algorithm}

\subsubsection{Termination} 
Any D2D UE can send a {\it Termination Request} message to the eNB so that the eNB terminates opportunistic relaying (see Fig.~\ref{fig:proto}). Next, the eNB will notify the termination of the communication by sending the {\it Termination Notification} message to the ProSe Function which then forwards this message to the ProSe Application Server.

\subsection{ProSe Amendments}
\label{ss:ProSeamend}
The above description of DORE procedures 
complies with the 3GPP's ProSe proposed architecture and procedures~\cite{3GPP23.703, 3GPP23.303}. However, we opt for a few modifications that improve system performance and security and reduce the relaying overhead.

{\bf Relay selection at eNB.} 
According to ProSe, the relay selection function should be implemented at the ProSe Server. However, such an implementation will result in additional delay. This can render the relay selection ineffective, as the channel quality of the relay may have changed until the decision of the ProSe Server is received and put into effect. To this aim, we propose to virtualize the relay selection function of ProSe to run our low overhead relay selection mechanism, which is simply choosing the best between two UEs, at the eNB.

{\bf Label switching instead of IP routing.}
The current relay solution in 3GPP uses IP routing to relay the traffic between UEs. Such an IP-based approach has a few caveats: $(i)$ The relay has to process LTE frames up to IP layer  to perform IP routing. This imposes {\it extra overhead to the relay} UE because the relay is subject to processes such as decompression and deciphering on behalf of the D2D receiver; $(ii)$ Deciphering the relay frames exposes the D2D receiver to {\it security threats} because the relay UE can potentially monitor the traffic at IP layer; and $(iii)$ The system should handle {\it IP mobility} because the D2D UEs have two IP addresses (i.e., one cellular link and another for D2D link) and the cellular data can be destined to either interface. We propose to label the packets by the eNB before sending them to the relay so that the relay knows which packets should be relayed over WiFi. Next, the relay encapsulates the PDCP packets in WiFi frames and transmits them to the destination. With this method, the IP handling issues can be disregarded. 

{\bf D2D link reporting.} Current 3GPP standard and academic literature assume that the capacity of WiFi is always higher than the cellular capacity~\cite{asadi2014ComCom,asadi2013WD}. This is a strong assumption, particularly in dense scenarios as WiFi operates on the unlicensed band that is used by various devices/technologies. 
To avoid overloading the relay UE beyond the capacity of the D2D link, we include an additional message to report average delay and capacity of the D2D link. 
This is crucial to keep QoS figures under control in the network.

 \begin{figure} [t!]
\centering
		\includegraphics[scale=0.4, angle=0]{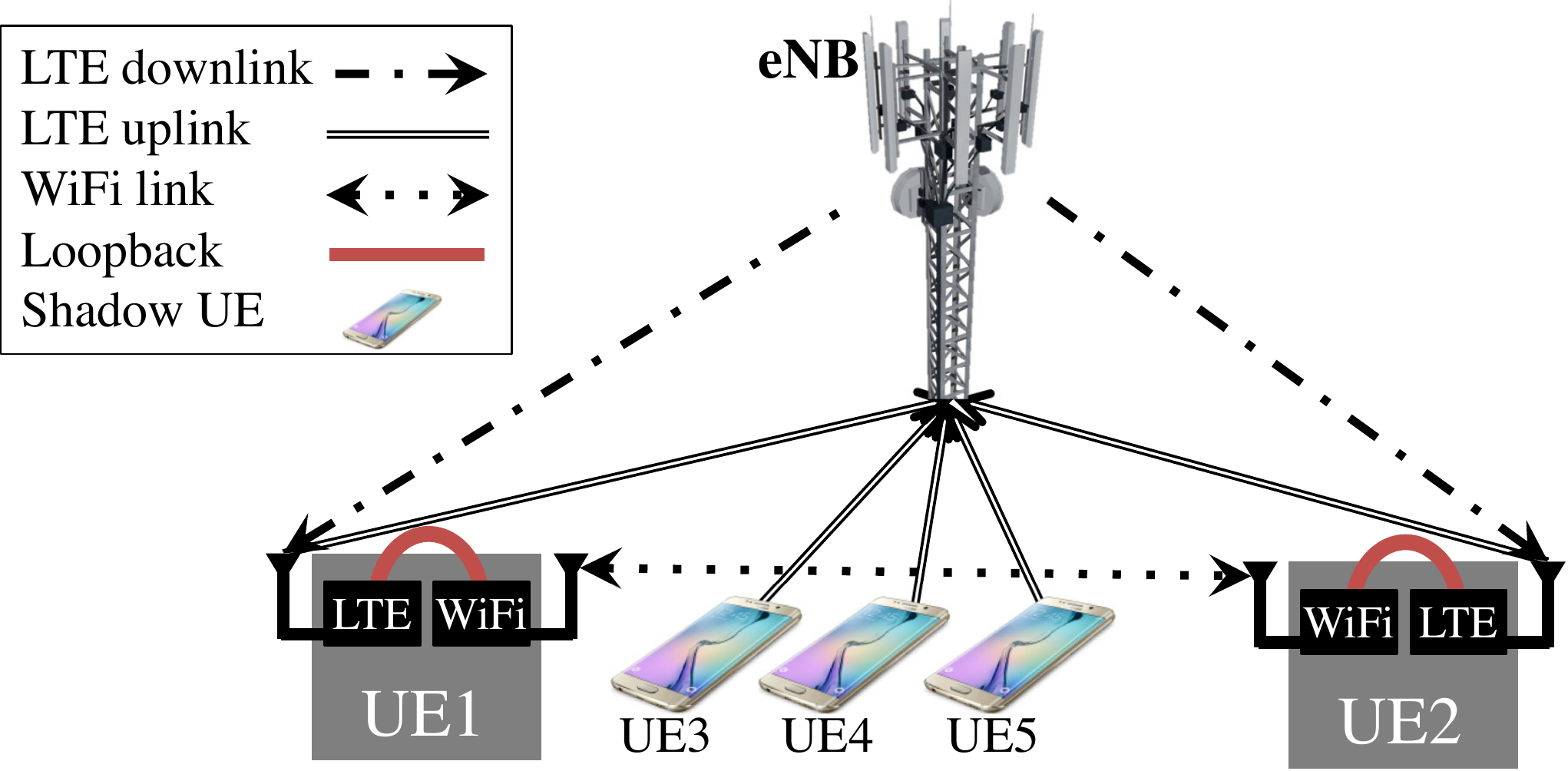} 
		\caption{Architecture of the testbed and the general setup of the experiments. In this setup, UE1 and UE2 are real UEs implemented in our SDR hardware. UE3 to UE5 are the so-called {\it shadow UEs}, i.e., off-the-shelf Android smartphones that simply provide their CQI to the eNB.}
		\label{fig:testbed}
\end{figure}

\section{Design and Implementation of the testbed}
\label{s:setup}
This section provides a detailed walk-through of our D2D implementation. As illustrated in Fig.~\ref{fig:testbed}, our testbed consists of three main components, namely, the eNB, the UEs and the {\it shadow UEs}. In what follows, we explain each component, its architecture and the interworking among different components.

\subsection{Software and Hardware}
We use LabVIEW\footnote{http://www.ni.com/labview/} SDR platform because it allows for quick implementation of CPU intensive physical layer operations with nano-second runtime requirement (e.g., Fast Fourier Transform (FFT), inverse FFT (iFFT), and coding) in a Xilinx FPGA. Moreover, it provides the means for high-speed communication with CPU/RF hardware. 

The required hardware for each UE/eNB is emboxed in an NI PXI 1082 chassis\footnote{http://sine.ni.com/nips/cds/view/p/lang/en/nid/207346} that contains: $(i)$ NI PXIe 8135 Real-Time controller\footnote{http://sine.ni.com/nips/cds/view/p/lang/en/nid/210545} operating on an Intel Core-i7-3610QE CPU. This controller hosts LabVIEW Real-Time OS that executes MAC and physical layer control algorithms with micro-second resolution; $(ii)$ NI FlexRIO module\footnote{http://www.ni.com/flexrio/} with Xilinx Kintex 7/Virtex 5 FPGA, which executes physical layer operations; and $(iii)$ NI 5791 FlexRIO Adaptor Module (FAM) that is used as an RF transceiver operating with a $100$ MHz bandwidth in the  frequency range from $200$ MHz to $4.4$ GHz. 
FAM is mainly used for Digital to Analog Conversion~(DAC) and Analog to Digital Conversion (ADC).

\subsection{Architecture of eNB}
The eNB consists of a Real-Time controller, a Virtex 5 FlexRIO, and a FAM for over the air LTE transmissions. 

{\bf Design.} Fig.~\ref{fig:eNB} shows the important blocks of the eNB. The Real-Time controller runs MAC layer operations such as scheduling, D2D services, and transport block generation for Control Channel (CCH) and shared channel (SCH). The FPGA executes physical layer operations such as interleaving for CCH traffic and scrambling for SCH traffic. Finally, the baseband signal is up-converted in the FAM module and transmitted over the air to the UE.  Moreover, we implemented Round Robin (RR) and Proportional Fair (PF)~\cite{margolies2014Infocom} schedulers at the eNB. The former is a benchmark commonly used in the literature. Both schedulers are used in today's cellular networks.

{\bf Communication.} The current testbed only supports OFDMA in downlink, and the uplink transmission is performed over Ethernet. However, in the future, we intend to extend this testbed to support OFDMA uplink transmission.
\begin{figure} [t!]
\centering
		\includegraphics[width=\columnwidth, angle=0]{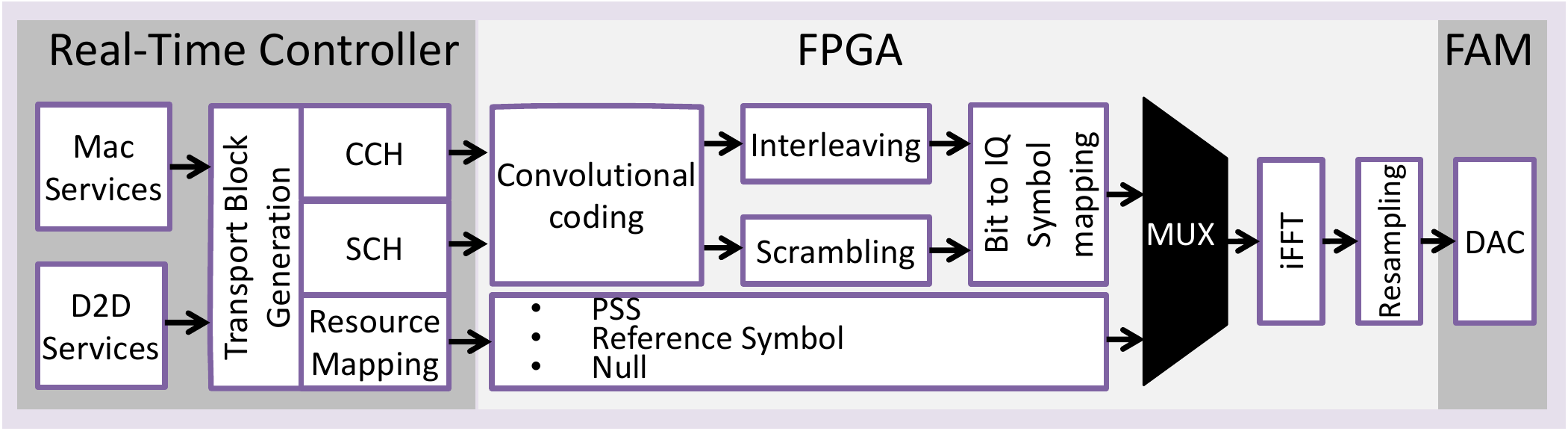} 
		\caption{Architecture of the eNB. The figure illustrates the location of each module described and the data flow among them.}
		\label{fig:eNB}
\end{figure}

\subsection{Architecture of the UE}
The UE consists of a Real-Time controller, a Virtex 5 FlexRIO as OFDMA receiver, a Kintex 7 FlexRio as WiFi transceiver, and two FAMs for over the air communications. 

{\bf OFDMA receiver.} 
As shown in Fig.~\ref{fig:UE-lte}, the DSP operations are implemented in the FPGA and the Real-Time controller handles the payload processing and MAC layer D2D operations. These operations consist in filtering the relay packets and transmitting them to the receiver over WiFi. 

\begin{figure} [h!]
\centering
		\includegraphics[width=0.95\columnwidth, angle=0]{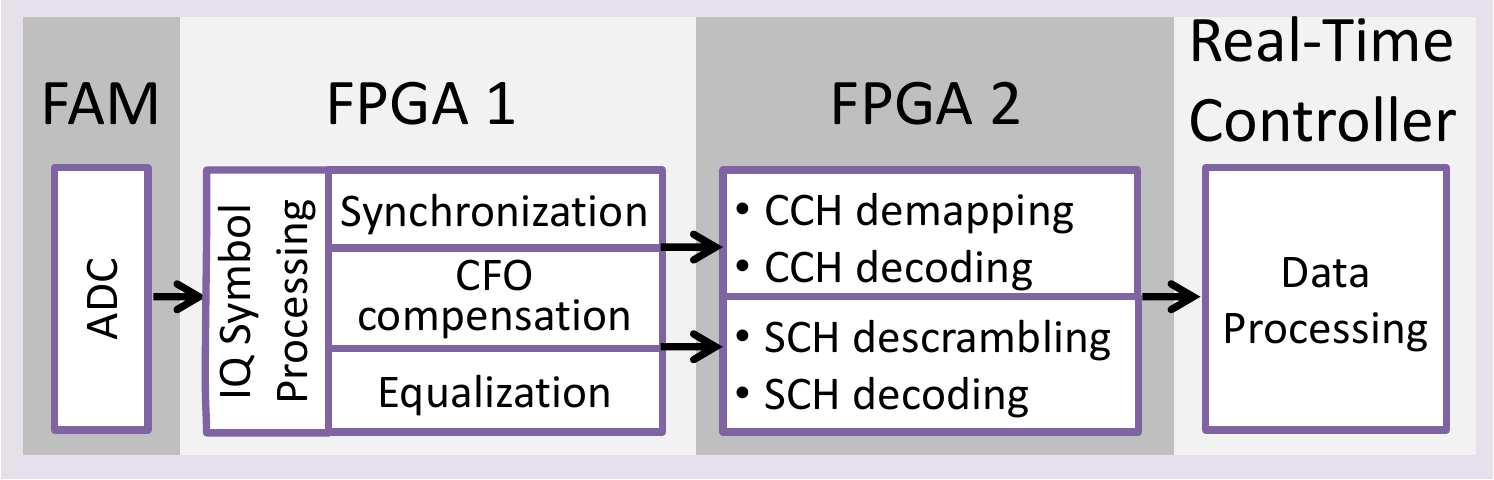} 
		\caption{Architecture of UE's LTE interface. The figure shows different modules and internal data flow.}
		\label{fig:UE-lte}
\end{figure}

{\bf WiFi transceiver.} The majority of the WiFi framework~\cite{wifi2014NI} is implemented in the FPGA, see Fig.~\ref{fig:UE-wifi}. In addition, the transceiver is implemented within the same FPGA. We implemented the D2D state-machine and its corresponding logic in the Real-Time controller. The controller is also in charge of feeding data to the FPGA transmission processing chain and reading the decoded data from FPGA processing chain. 

\begin{figure} [h!]
\centering
		\includegraphics[width=0.85\columnwidth, angle=0]{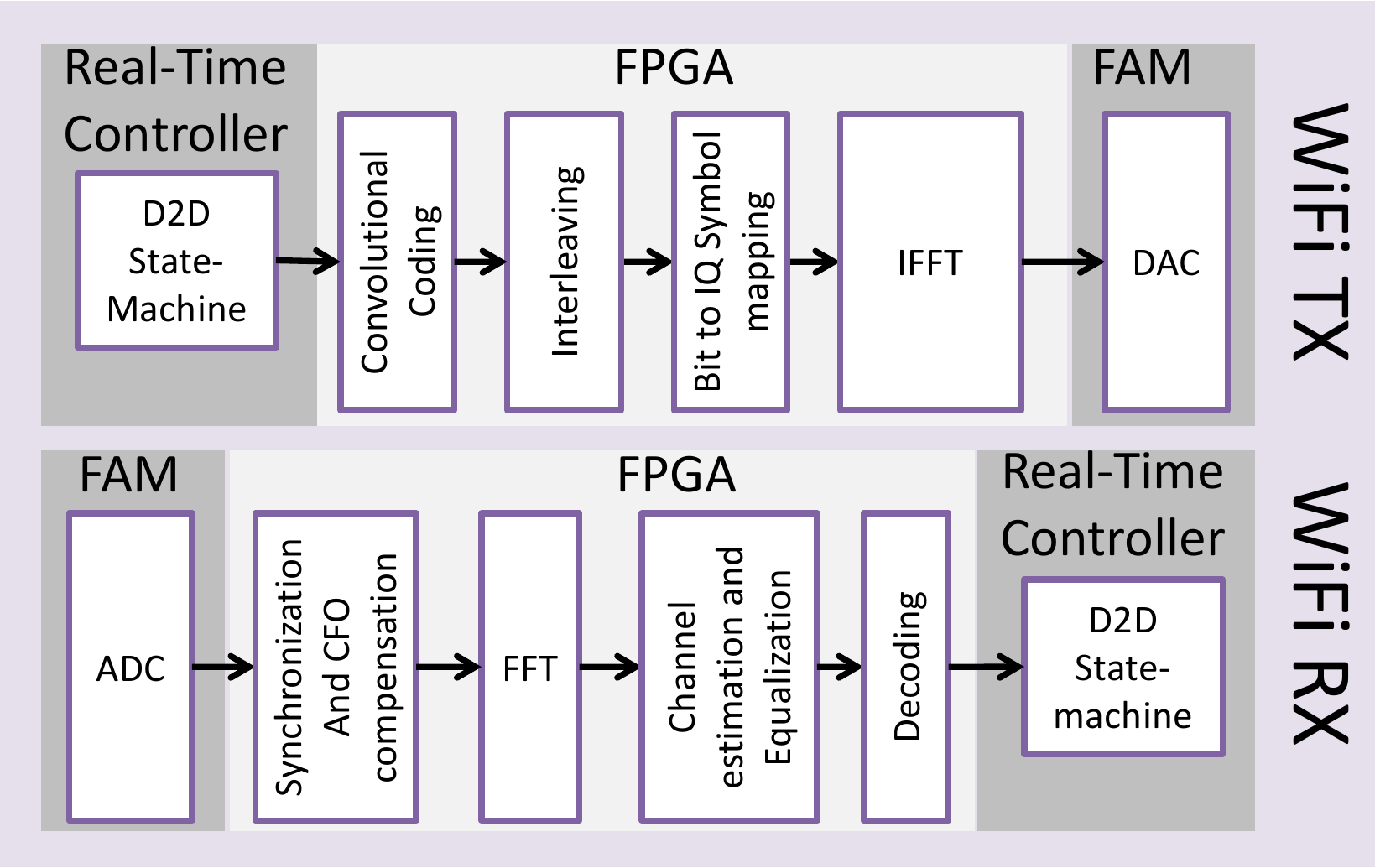} 
		\caption{Architecture of UE's WiFi interface. The figure illustrates the location of each of the modules described and the data flow among them.}
		\label{fig:UE-wifi}
\end{figure}

{\bf Communication.} We observe in Fig.~\ref{fig:testbed} that UEs receive downlink transmissions from the OFDMA receiver and send the uplink messages over an Ethernet link. The WiFi (i.e., D2D) communication uses an OFDM transceiver. 
\subsection{Shadow UEs}
\label{ss:shadow}
These UEs (i.e., UE3, UE4, and UE5 in Fig.~\ref{fig:testbed}) are  off-the-shelf Android smartphones. We include the shadow UEs in our setup to better capture the performance of outband D2D in a real-world scenario. We developed an Android application to obtain real-time cellular channel quality on a millisecond basis.
 The application then transmits the channel quality values to an access point, which is connected to the eNB over an Ethernet link. Although the shadows do not receive the actual transmission, the eNB schedules them and transmits their data as if they were real UEs. Since the mapping between MCS and Signal-to-Noise Ratio (SNR) is done such that the block error rate remains below $10^{-4}$, we assume that the shadow UEs receive the transmitted blocks with success probability of $0.9999$.

{\bf Communication.} Shadow UEs send their CQIs to a wireless access point which is connected to the eNB via Ethernet.

%

%
%
%
%

\subsection{Synthetic Fading}

Due to the limitation in the number of equipment in our disposal, we must run each experiment  at a separate time instant. In an ideal case, the system can be connected to high-end multi-channel cellular channel emulators to create the same channel variation in each experiment. Since we do not have such a device, we create a repeatable channel variation situation using refractors. 
In order to create repeatable channel variation patterns, we mounted the refractors plates on a step motor that is controlled by an Arduino Uno\footnote{http://www.arduino.cc/en/Main/ArduinoBoardUno} micro-controller. We generate synthetic channel variation by changing the rotation speed of the step motor. 
Fig.~\ref{fig:mcsCdf} is the proof-of-concept of this mechanism. We repeated an experiment four times and plotted the Cumulative Distribution Function (CDF) of the potential MCS (obtained from channel qualities) for both UEs to ensure stable repetitions of the channel variation. Indeed, the results show that this approach is suitable to re-create the same channel environment for different experiments.  Note that the location/frequency is selected such that unpredictable interference is minimized.

\begin{figure} [t!]
\centering
		\includegraphics[scale=0.53, angle=0]{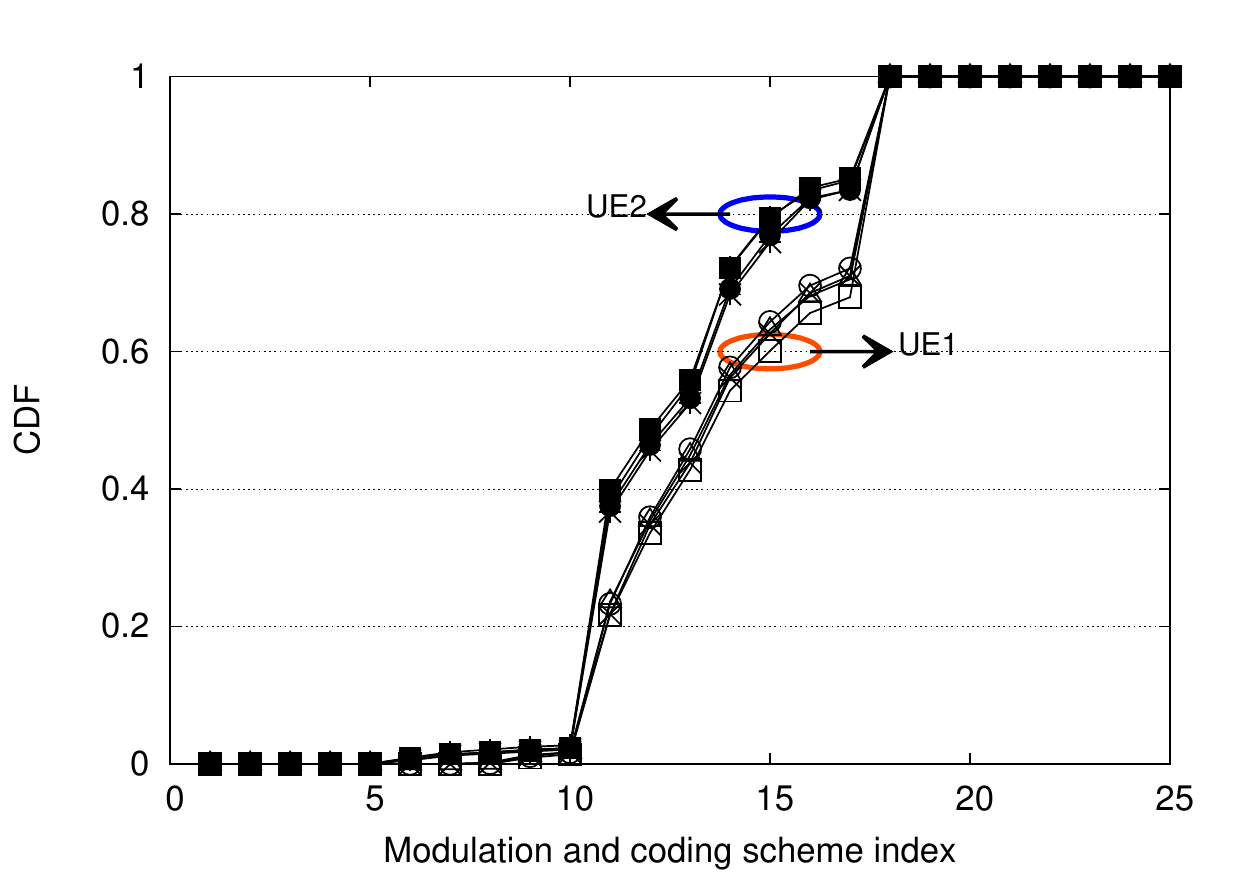} 
		\caption{CDF of MCS for UE1 and UE2 in each experiment. The figure shows that UE1 has on average a higher channel quality than UE2.}
		\label{fig:mcsCdf}
\end{figure}

\section{Experimental Evaluation}
\label{s:eval} 
In this section, we experimentally evaluate the performance of outband D2D-relay and DORE. 
We design several experiments to better demonstrate the system behavior in different scenarios. The general experiment setup is portrayed in Fig.\ref{fig:testbed}. 
We first present the performance of a simple outband D2D-relay setup. The simple setup is then redesigned to first incorporate channel opportunism and then QoS-awareness. We also examine the impact of non-collaborative UEs (i.e., shadow UEs).
The duration of each experiment is $300$ s, which is sufficiently long to observe the average system's performance. 
%
In order to provide the reader with a detailed view of the achieved performance, we show minimum, maximum, $25^{th}$ and $75^{th}$ percentiles in addition to the average values. Unless otherwise specified, the rotation speed of the refractor is $5$ rpm. Finally, UE1 experiences higher average channel quality than UE2 in all experiments.

\subsection{Selected KPIs}

We report several KPIs to examine different aspects of outband D2D-relay and DORE. The KPIs described below are chosen based on their importance for understanding the characteristics of a practical D2D system. 

{\bf Throughput.} Throughput is measured as the number of received bits per second. 

{\bf Delay.} We timestamp each packet at the eNB MAC and measure the delay at three points within the path from the eNB to the D2D receiver.{\it LTE delay} refers to the time taken for a packet to arrive from the eNB's MAC layer to the MAC layer of the UE on the LTE interface. The {\it cross-platform delay} measures the time taken to send a packet from MAC layer in LTE stack to the same layer at the WiFi stack. This value basically highlights the overhead imposed by processing/transferring data between two interfaces at the same UE. {\it WiFi delay} is the delay experienced by a packet to reach from the WiFi MAC layer at the relay to the D2D receiver's WiFi MAC. Finally, the {\it end-to-end delay} is the sum of all the above described delays.



{\bf CPU load.} Since the Real-Time controller executes the D2D related operations, we can provide the extra CPU load due to D2D operations by monitoring the Real-Time module. 

{\bf D2D lifetime.} We examine our proposed design with slow and fast channel variations. In each case, we measure the time during which a UE acts as a relay, which we call {\it relay lifetime}. This is a major factor in opportunistic D2D because frequent  role switching imposes extra load to the system. 

{\bf Structural Similarity.}  This is an index of similarity between two images, and it is known to be a better estimation of  human eye perception in comparison to other traditional methods such as peak SNR or mean squared error. We use this metric for QoE measurements in video streaming experiments.

\begin{figure*}[!t]
	\centering
	\def\subfigcapskip{-3pt}
		\subfigure[Delay at different parts of the path.]
		{
			\includegraphics[width=0.62\columnwidth]{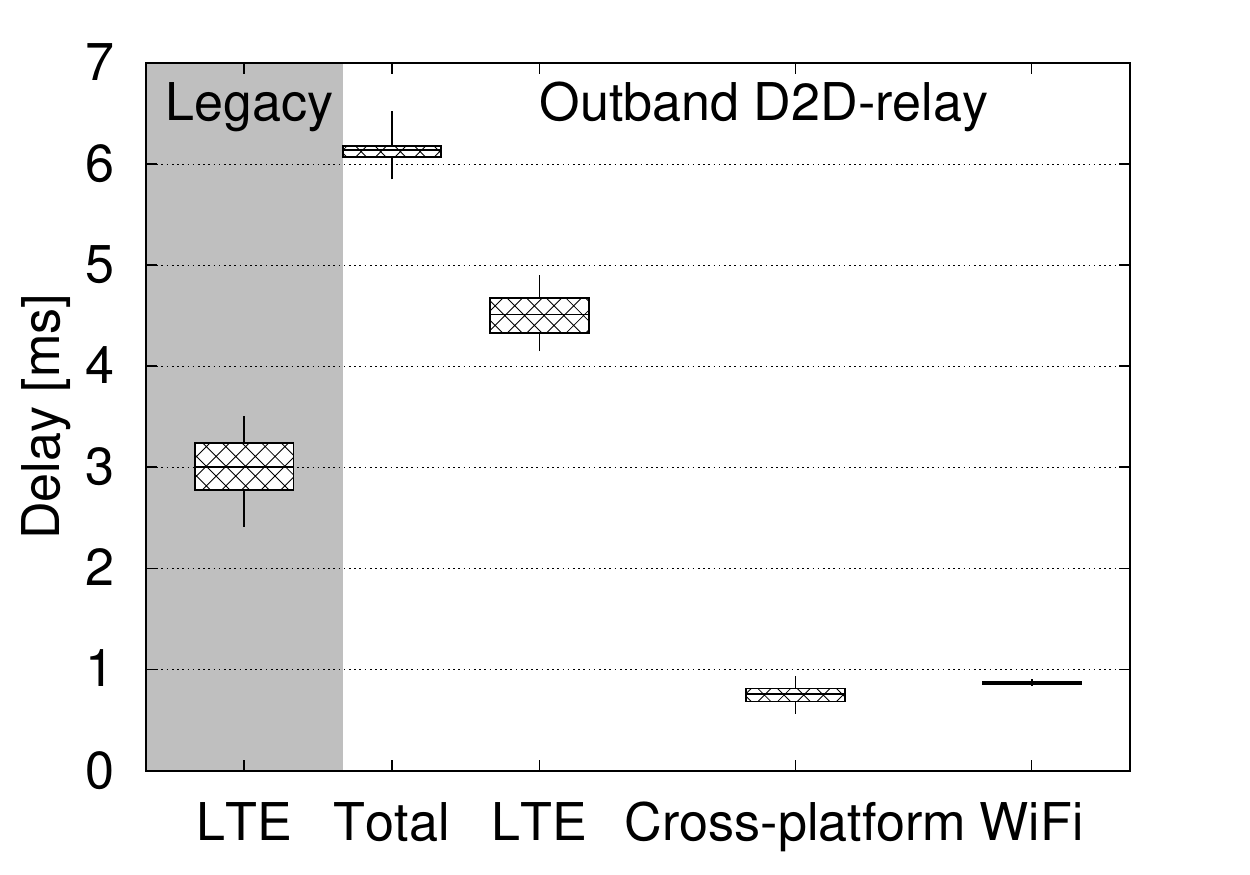} 
			\label{fig:simpDelay}
		}
		\subfigure[Per-UE throughput.]
		{
			\includegraphics[width=0.62\columnwidth]{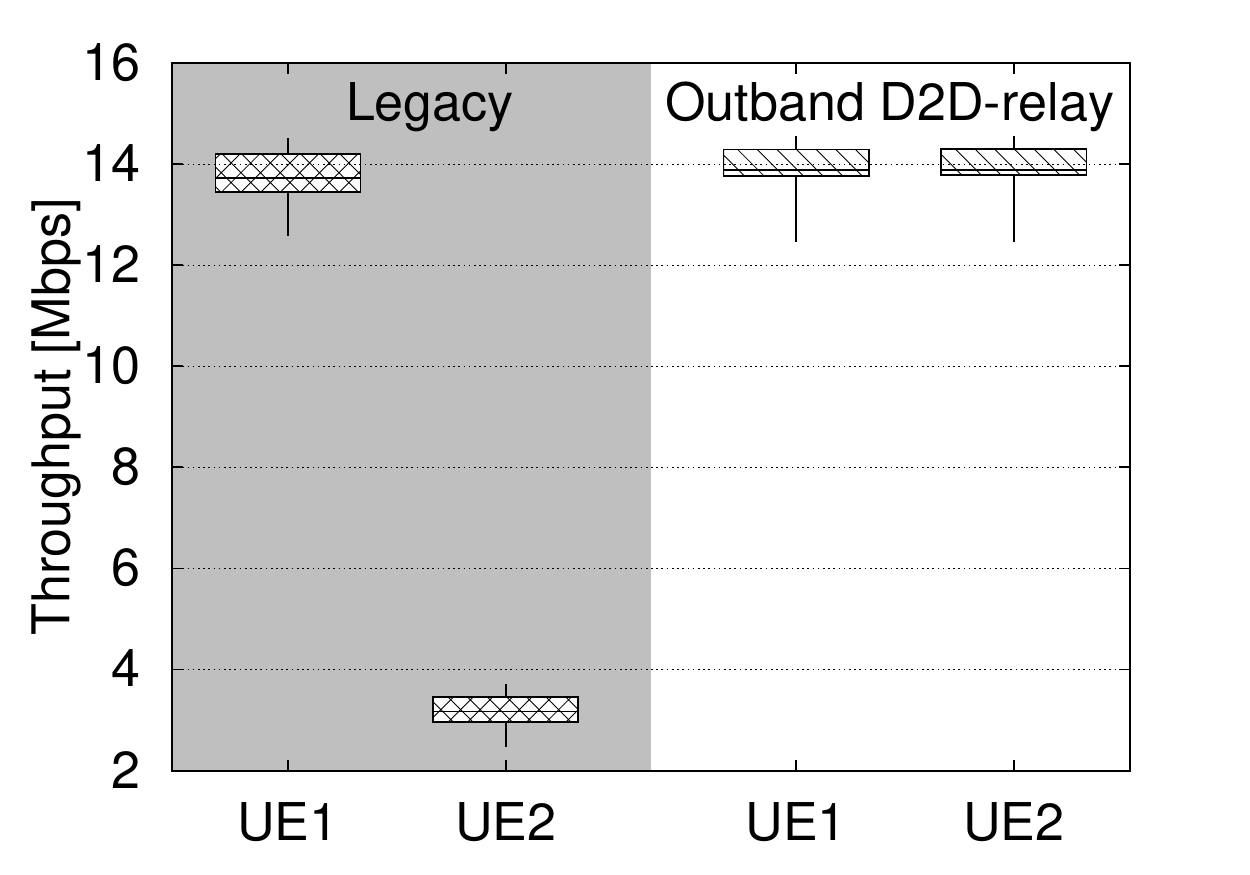} 
			\label{fig:simpTput}
		}
		\subfigure[CPU load.]
		{
			\includegraphics[width=0.64\columnwidth]{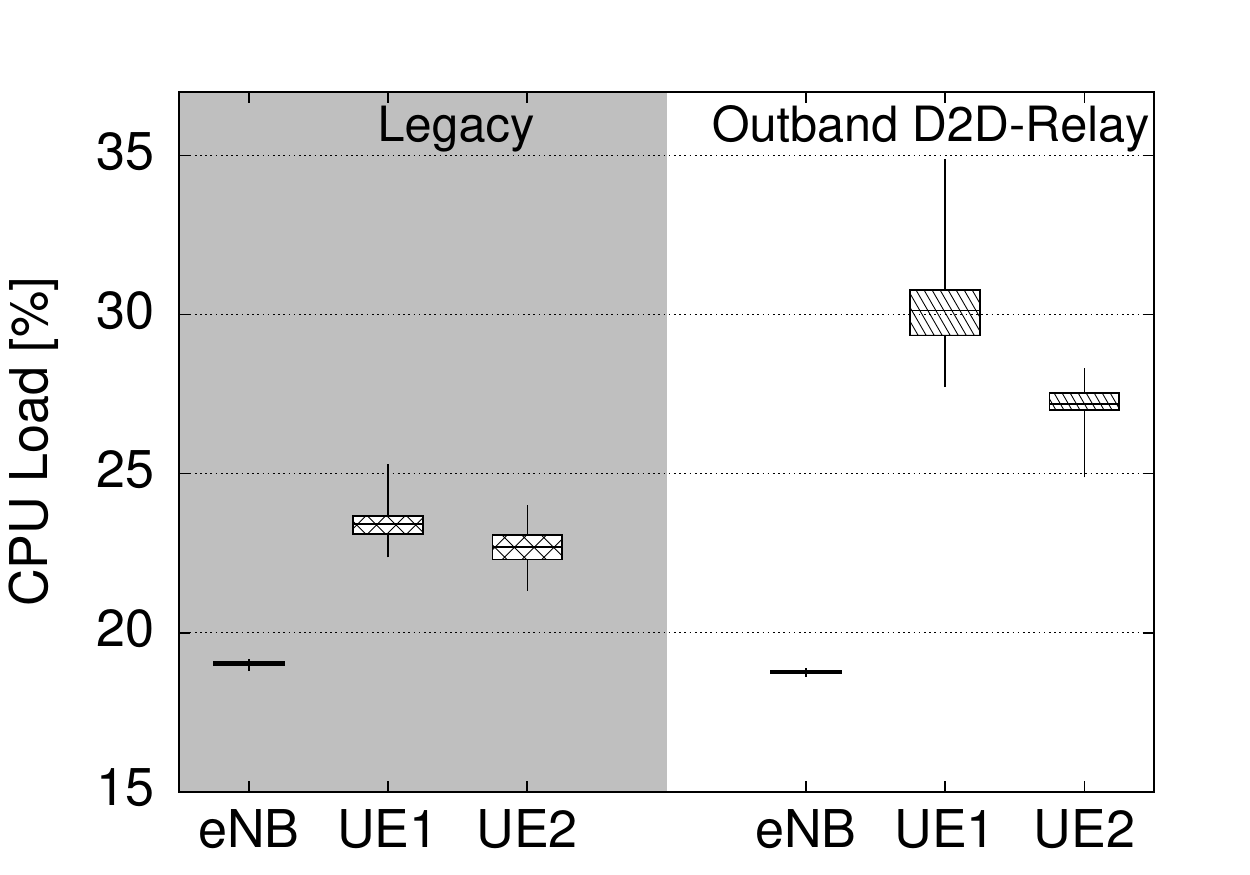} 
			\label{fig:simpLoad}
		}

		\caption{Outband UE-Relay: UE1 relays the traffic from the eNB to UE2}
		\label{fig:simple}
\end{figure*}

\subsection{Non-opportunistic Outband D2D Relay}
We start with the simplest form of outband D2D-relay scenario with two UEs. 
Despite the simplicity of this experiment, it provides answers regarding the delay overhead due to multi-hop communication and achievable throughput gain.

Fig.~\ref{fig:simpDelay} compares a {\it Legacy} scheme (in which both UEs receive traffic only from the eNB) with an {\it Outband D2D-relay}, in which UE1 acts as a relay for UE2. We observe that outband D2D increases the average end-to-end delay (i.e., Total in the figure) by $3.3$~ms as compared to the Legacy cellular system. 
Looking at different delay components of outband D2D-relay, we can see that cross-platform delay and WiFi delay are the major contributors to the delay overhead ($2$ ms out of the $3.3$ ms total delay overhead). It is important to note that cross-platform delay caused by extra frame processing at the relay, which results in higher LTE delays in outband relay mode. While commonly ignored in the literature, this illustrates that relaying large volumes of traffic comes at a cost. According to the observation from the delay profile, outband relay could be potentially suitable for a large variety of non-mission critical applications. Indeed, outband relay with a total delay of $6.3$~ms meets the 3GPP suggested delay budget of $70$~ms~\cite{3GPP23.203}. 

The motive for opportunistic D2D-relay is vividly depicted in Fig.~\ref{fig:simpTput}. The figure shows that UE2 suffers from low channel quality while UE1 experiences a good channel condition. After outband D2D activation, UE2's throughput increases significantly because it receives its traffic through a high channel quality relay. 


We measured the CPU load of each device with and without outband D2D. Our observations in Fig.~\ref{fig:simpLoad} show that UE1 (i.e., relay) and UE2 (D2D receiver) are subject to $6.3$\% and $4.2$\% CPU load overhead because of outband D2D operations. The overhead is negligible at the eNB. Note that running the WiFi code in the idle mode on the Real-Time controller increases the total CPU load by about $4$\%. Hence, if we assume that the UEs WiFi interface is in the idle mode, the overhead due to outband D2D is marginal.

\subsection{DORE with Delay-tolerant Traffic}
Now, we evaluate the performance of opportunistic outband D2D using RR and PF scheduling algorithms. 
We test DORE with delay-tolerant traffic (i.e., no delay threshold in Algorithm I)
to evaluate the potential throughput gain for such use-cases.
In the figures, we label the legacy schemes as RR and PF. When used for DORE with delay-tolerant traffic, they are labeled as  RR-DT and PF-DT.  
\begin{figure*}[!t]
	\centering
	\def\subfigcapskip{-3pt}
		\subfigure[Aggregate throughput.]
		{
			\includegraphics[scale=0.45]{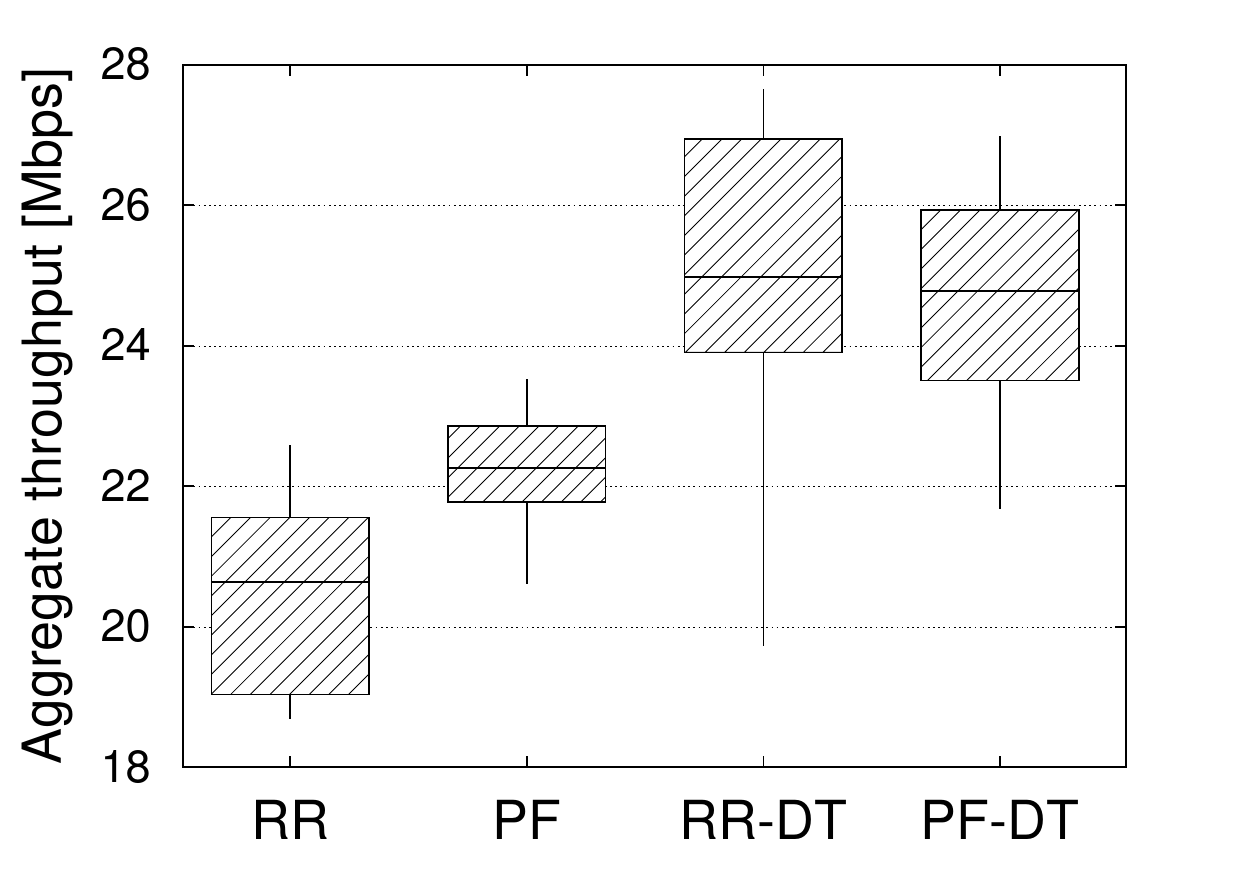} 
			\label{fig:2usrTput}
		}
		\subfigure[Delay.]
		{
			\includegraphics[scale=0.45]{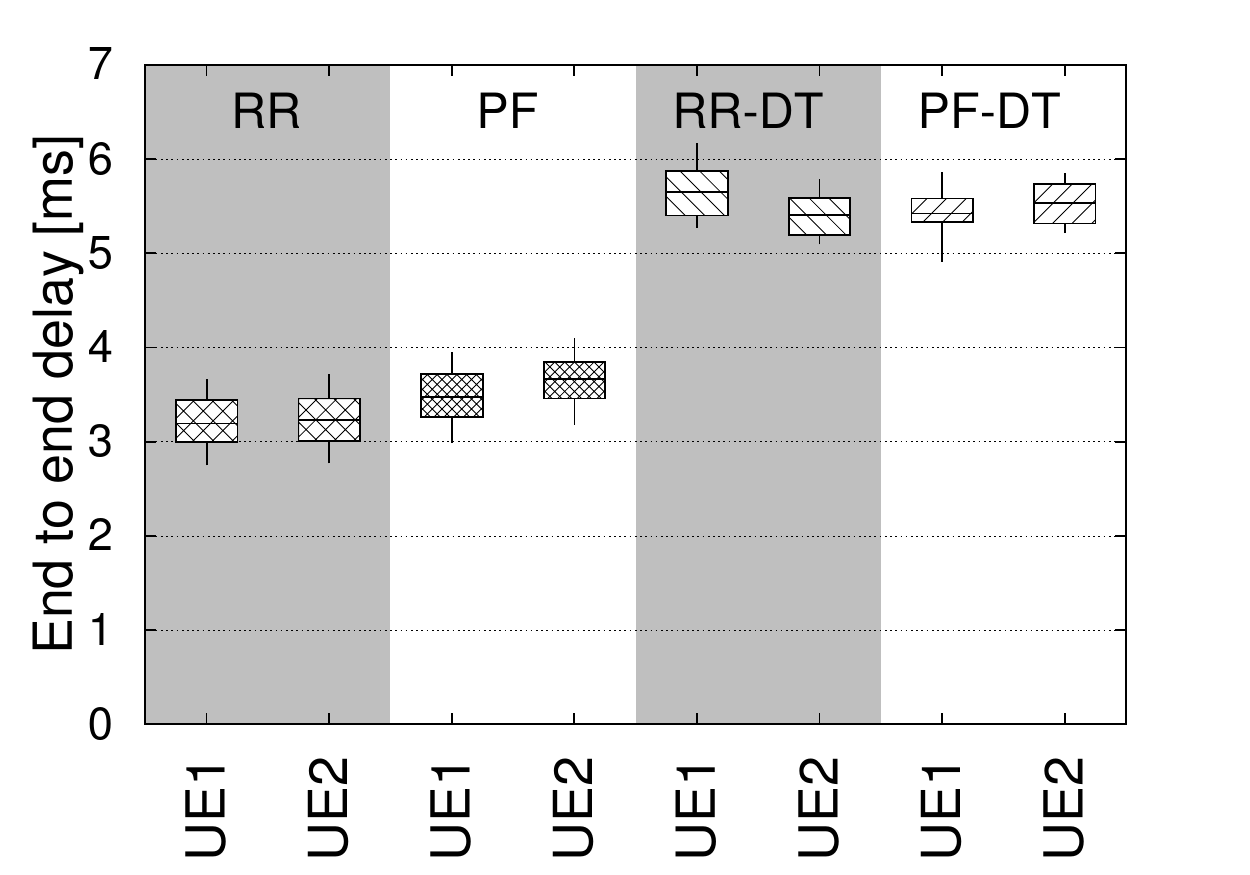} 
			\label{fig:2usrDelay}
		}
		\subfigure[CPU load.]
		{
			\includegraphics[scale=0.45]{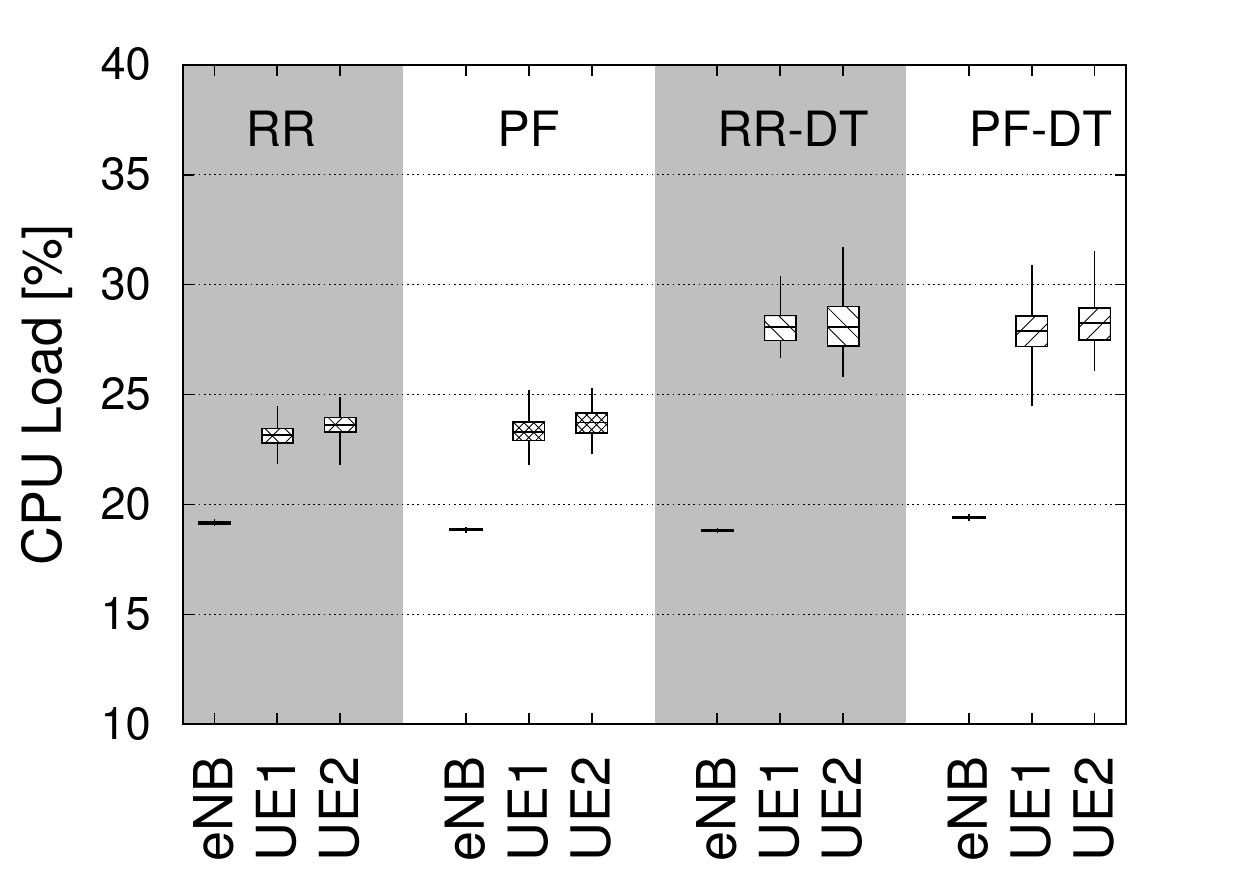} 
			\label{fig:2usrLoad}
		}		
		\caption{DORE: the relay UE is chosen according to reported CQI values.}
		\label{fig:twouser-1}
\end{figure*}

Fig.~\ref{fig:2usrTput} shows the achievable aggregate throughput of RR-DT and PF-DT is $21$\% and $11.2$\% higher than RR and PF, respectively. As mentioned in Section~\ref{s:proto}, opportunistic outband D2D leverages the channel diversity between the D2D users. Since PF harvests part of this opportunism due to its opportunistic nature, the resulting gain reduces by $9.8$\% in comparison to RR. Nevertheless, the gain remains relevant for a two-user scenario where there are limited opportunities. We show later in this section that the opportunistic gain increases with the user population. Delay comparison in Fig.~\ref{fig:2usrDelay} demonstrates DORE causes higher delays. The additional delay stems from WiFi and cross-platform transmission and LTE frame processing. 



Fig.~\ref{fig:2usrLoad} depicts the impact of opportunistic outband D2D on CPU loads in the eNB and the UEs. The impact of opportunistic outband D2D is negligible on eNB. On the other hand, the UEs experience about $6$\% additional CPU load that is mainly due to WiFi operations. Unlike our observation in the previous non-opportunistic scenario (Fig.~\ref{fig:simpLoad}), UE1 and UE2 have similar CPU load. This is due to the fact that in opportunistic outband D2D, the relay changes dynamically based on the reported CQI. Thus, both UEs act as a relay in a portion of the time, allowing load balancing/sharing between the UEs. In an extreme case where one of the UEs always has the lowest channel quality, we will observe similar results as shown in Fig.~\ref{fig:simpLoad}.

\subsection{Impact of Fading Speed}

\begin{figure*}[!t]
	\centering
	\def\subfigcapskip{-3pt}
		\subfigure[Impact of refractor speeds on D2D lifetime.]
		{
			\includegraphics[scale=0.45, angle=0]{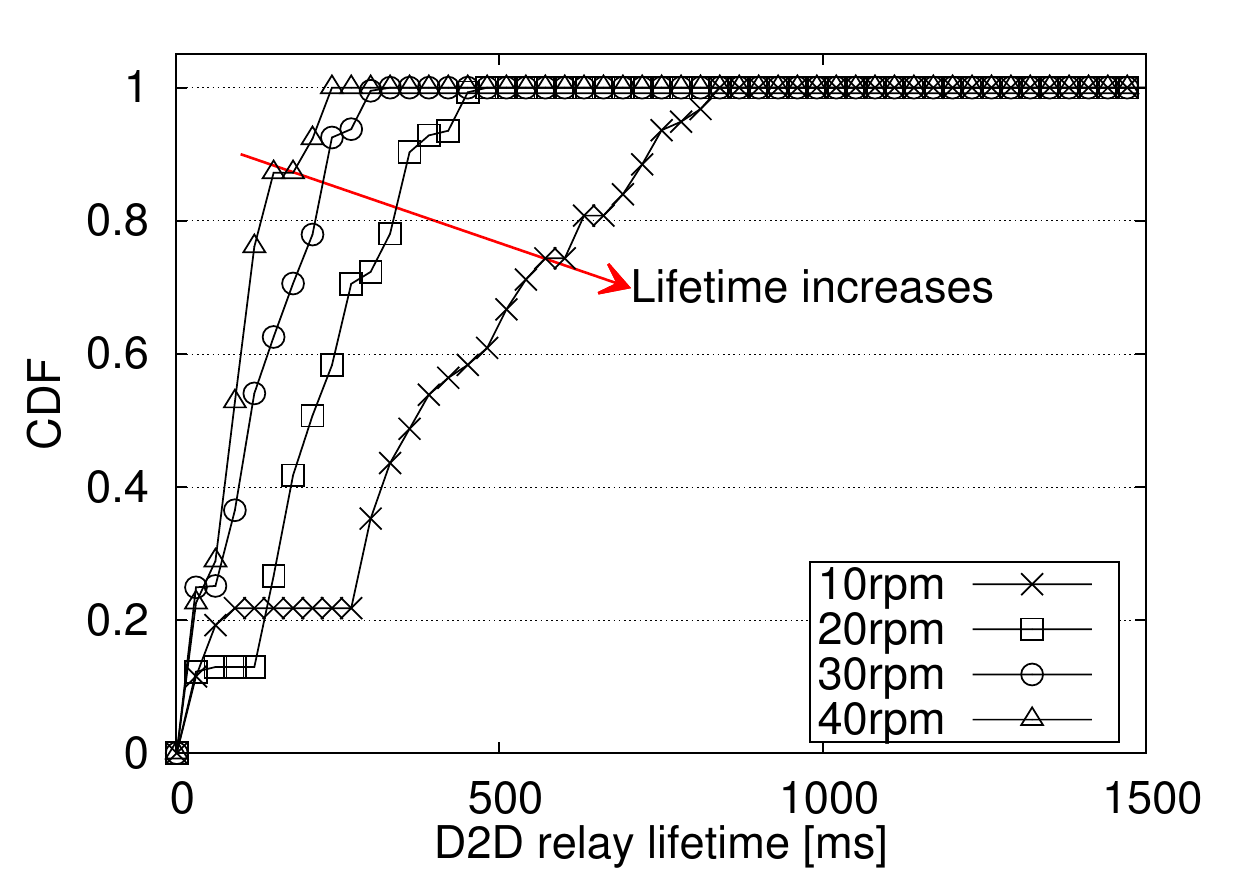} 
			\label{fig:fadingLifecycle}
		}
		\subfigure[Impact of  threshold on D2D lifetime.]
		{
			\includegraphics[scale=0.45, angle=0]{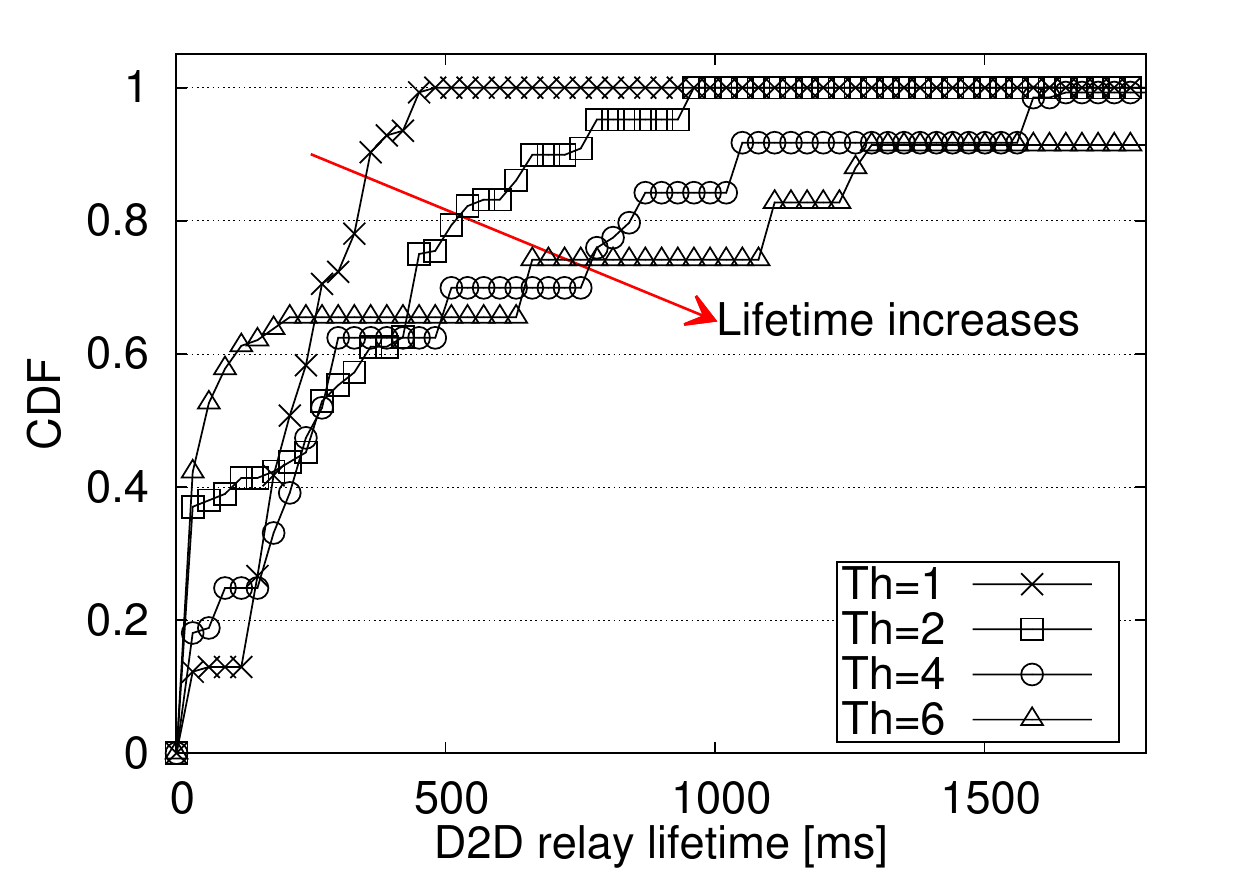} 
			\label{fig:fadingThresh}
		}
		\subfigure[Impact of switching thresholds.]
		{
			\includegraphics[scale=0.45, angle=0]{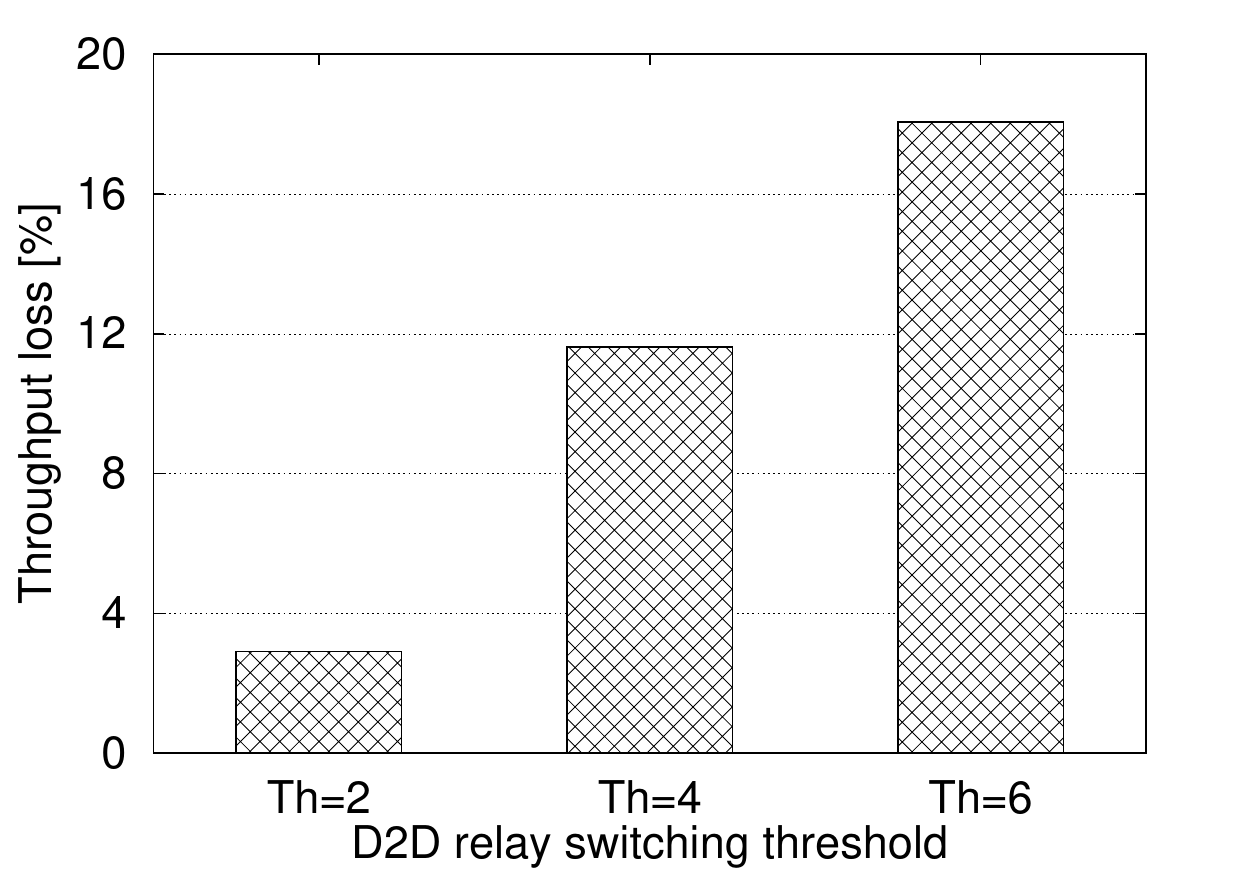} 
			\label{fig:fadingTput}

		}
		\caption{Impact of fading speed on the lifetime of D2D UEs.}
		\label{fig:fading}
\end{figure*}

\begin{figure*}[!t]
	\centering
	\def\subfigcapskip{-3pt}
		\subfigure[Per-UE throughput.]
		{
			\includegraphics[scale=0.45, angle=0]{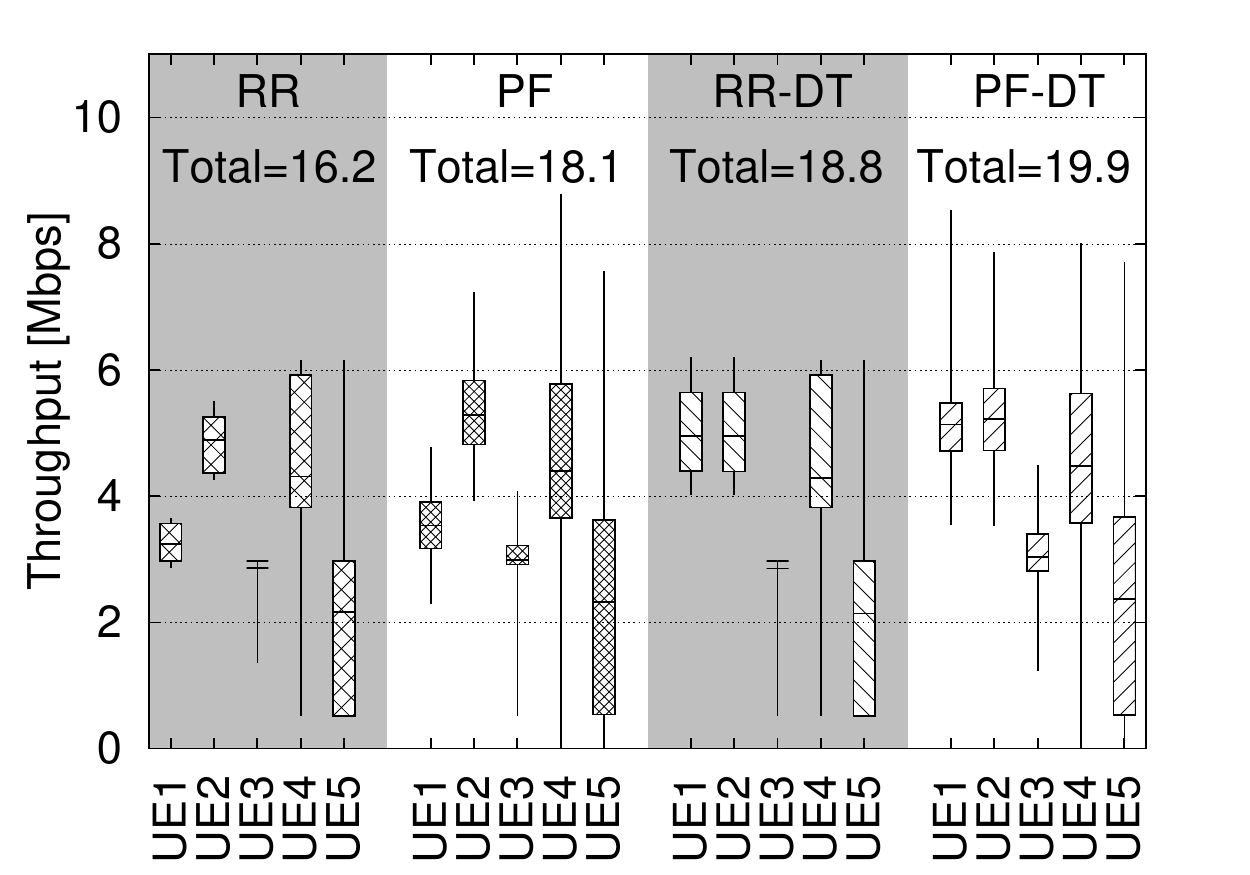} 
			\label{fig:shadowTput}
		}
		\subfigure[End to end delay for D2D UEs.]
		{
			\includegraphics[scale=0.45, angle=0]{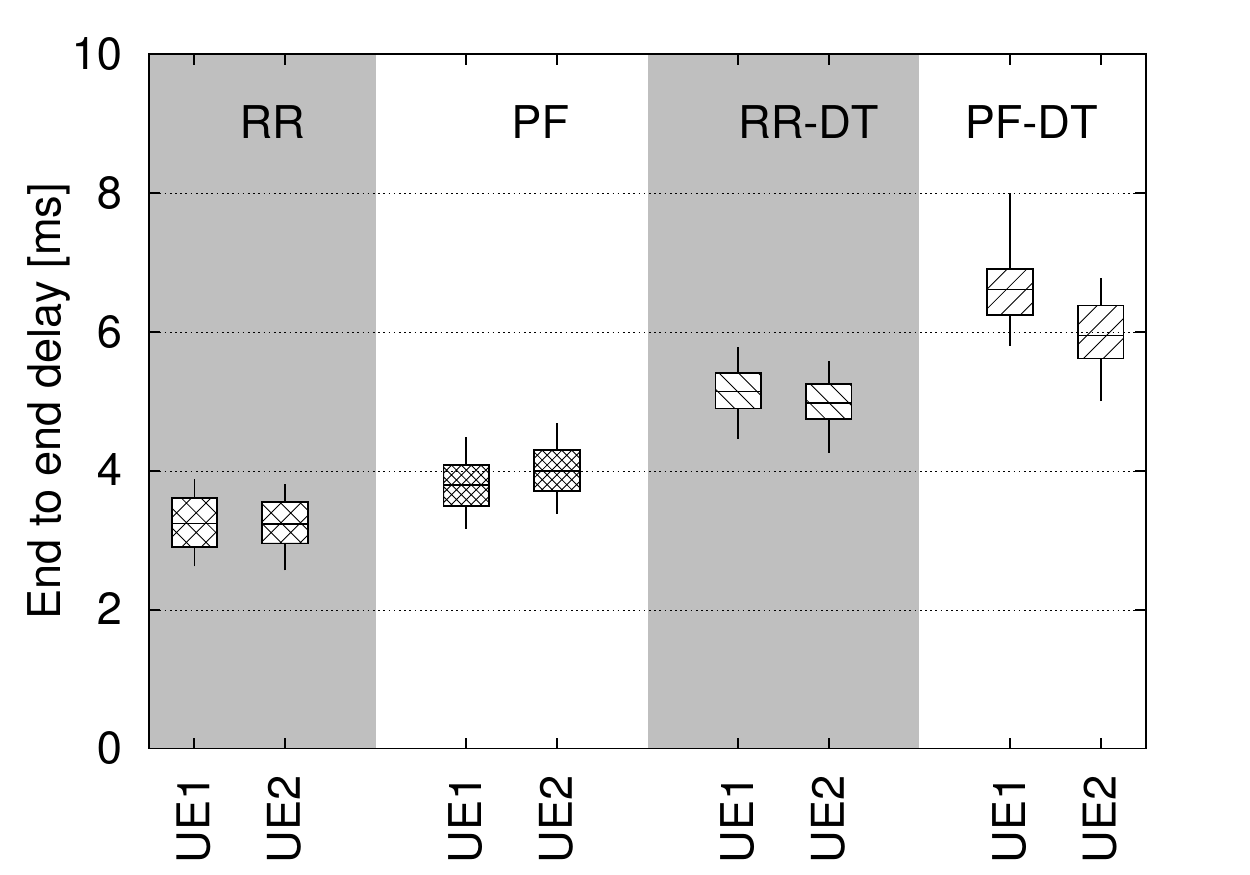} 
			\label{fig:shadowDelay}
		}
		\subfigure[CPU load.]
		{
			\includegraphics[scale=0.45, angle=0]{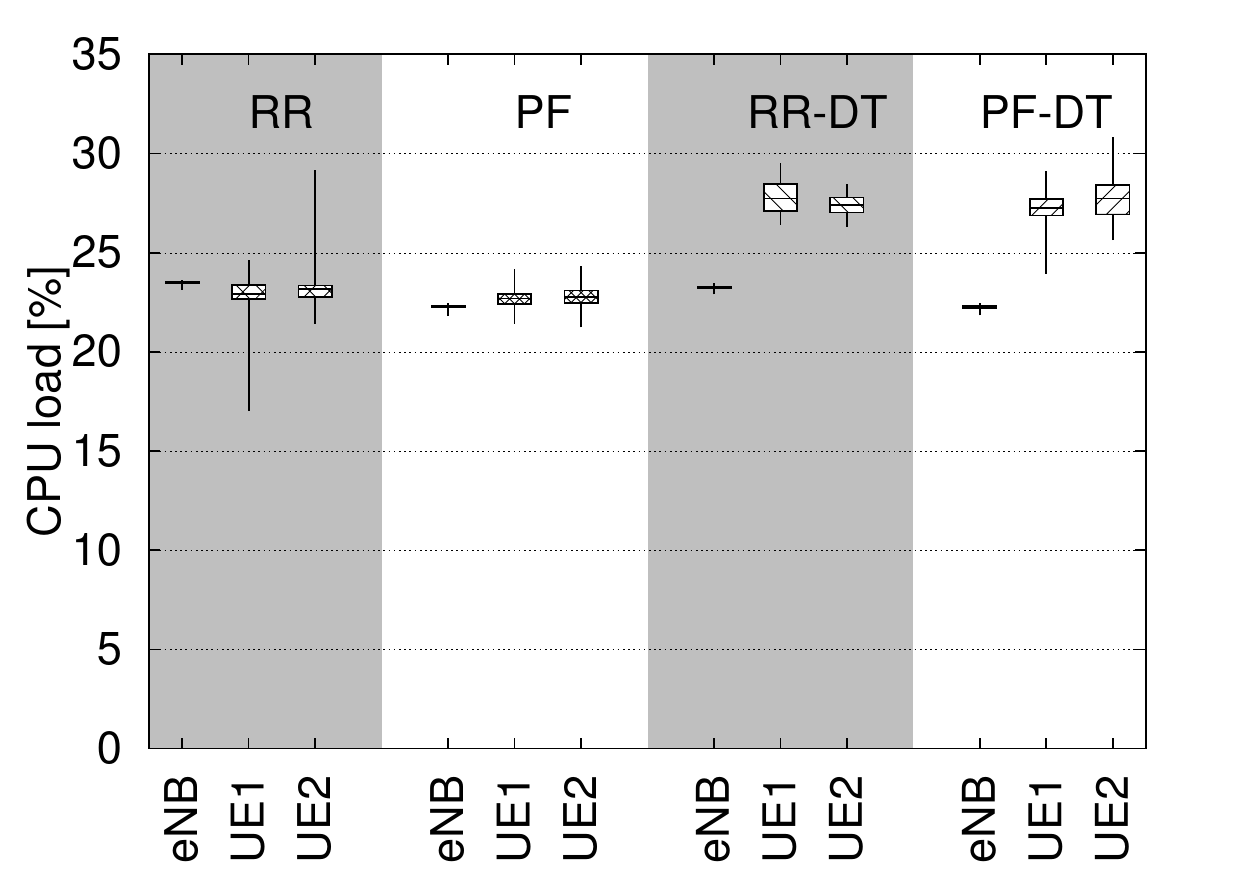} 
			\label{fig:shadowCPU}
		}
		\caption{System KPIs in an experiment with two D2D UEs and three shadow users.}
		\label{fig:shadow-1}
\end{figure*}

This experiment is designed to show the dynamics of DORE under different fading scenarios. 
In particular, the change of role in the D2D connection (i.e., a UE can be a relay or a D2D receiver). We refer to the period in which a D2D UE acts as a relay as the {\it lifetime}. In this experiment, we shed light on the frequency of these changes and their impact on the system.

Fig.~\ref{fig:fadingLifecycle} shows the CDF of the lifetime of UEs when the refractor surface spins at $10$, $20$, $30$, and $40$~rpm. At these rotation speeds, the MCS of a UE remains the same for $18.86$~ms,  $15.38$~ms, $13.51$~ms, and $10.82$~ms, on average. We can see that the duration of the lifetime increases as the fading speed reduces. The results also show that regardless of fading speed, the  lifetime is  shorter than $250$ ms more than $50$\% of the time. This emphasizes on the fact that {\it any implementation of opportunistic outband D2D must be capable of handling the relay dynamics on a millisecond timescale}.  

In our implementation of DORE, a switch of D2D roles occurs as soon as the achievable MCS of the D2D receiver becomes higher than the one of the relay UE. In other words, the MCS difference threshold to switch roles is one MCS index. Nevertheless, considering the resulting short lifetimes depicted in Fig.~\ref{fig:fadingLifecycle}, we have decided to introduce and test hysteresis in the switching to reduce frequent switching.  Introducing higher switching threshold can avoid role changes due to small MCS variations that do not vary much in terms of bit efficiency. Thus, we increase the MCS difference that triggers the role switching. 
Fig.~\ref{fig:fadingThresh} shows that larger thresholds (Th in the figure) increase lifetimes, as expected. However, this increment comes at the cost of reduced throughput. Indeed, Fig.~\ref{fig:fadingTput} illustrates that the throughput reduces up to $18$\% when the switching threshold is $6$ MCS levels.  Our results indicate that small switching thresholds increase D2D lifetime with limited throughput penalty. Therefore, it is not strictly necessary to reconfigure D2D links upon any MCS change, which reduces the complexity of the implementation.

\subsection{DORE in the Presence of Shadows and Delay-tolerant Traffic}

Here, we emulate the presence of additional legacy UEs using the shadow UEs introduced in Section~\ref{ss:shadow}. The shadows do not collaborate in DORE, but they help us to test DORE in the presence of non-collaborative UEs. 
The shadows send real-time CQI reports to the eNB, and the eNB schedules traffic for them, although they cannot decode such traffic.

Per-UE throughput results are presented in Fig.~\ref{fig:shadowTput}. We can see that UE1 achieves a $53.2$\% throughput gain with DORE (i.e., RR-DT and PF-DT) while UE2 only achieves a mere $1.4$\%  throughput gain.  UE2 achieves lower gain due to its higher average channel quality. 
We also reported the aggregate throughput of each scheme in Fig.~\ref{fig:shadowTput}, marked as {\it Total}. DORE results in $10.2$\% and $9$\% throughput gain compared to RR and PF. The throughput gains are lower than those achieved in the previous scenario ($\sim 20$\%). This is because in a scenario with $5$ UEs, the relay UE receives only a fraction of the total available bandwidth (i.e.,  $2/5$ of the resources can be relayed if RR is used). As a result, the opportunistic scheme can only optimize that portion of the cellular resources. 

Fig.~\ref{fig:shadowDelay} depicts the end-to-end delay. The delay behavior of the UEs is very similar to the delay behaviors observed in Fig.~\ref{fig:simpDelay}. Both UEs experience additional delay under RR-DT and PF-DT w.r.t. RR and PF because of the aforementioned cross-platform and WiFi delays.  UE1 has a higher delay than UE2 because it has lower channel quality than UE2 and it acts as the D2D receiver most of the time.

Fig.~\ref{fig:shadowCPU} compares the CPU load of the eNB and the UEs. The overhead on the eNB is negligible.
The two D2D-enabled UEs experience $4.42\%$ and $4.45\%$ higher CPU load due to outband D2D operations in WiFi and LTE interfaces. Note that running the WiFi code in the idle mode on the Real-Time controller increases the total CPU load by about $3$\%. Hence, 
the overhead due to outband D2D is marginal. 

\subsection{DORE in the Presence of Shadows and Delay-sensitive Traffic}
\label{ss:evalDore}
In this experiment, 
UE1 and UE2 host a real-time gaming application and a Voice over IP (VoIP) call with $30$ ms and $80$ ms over the air delay budget, respectively. To highlight the impact of  DORE's QoS-awareness, we also show the performance figures when the delay thresholds are set to infinity (i.e., DORE ignores the delay constraints). In this scenario, we stressed the WiFi channel (i.e., D2D link) by introducing extra non-D2D traffic to the network so that the WiFi channel operates near to the congestion point. Therefore, small changes in the instantaneous channel quality provoke non-negligible size queues. 
\begin{figure*}[!t]
	\centering
	\def\subfigcapskip{-4pt}
		\subfigure[Aggregate throughput.]
		{
			\includegraphics[scale=0.45, angle=0]{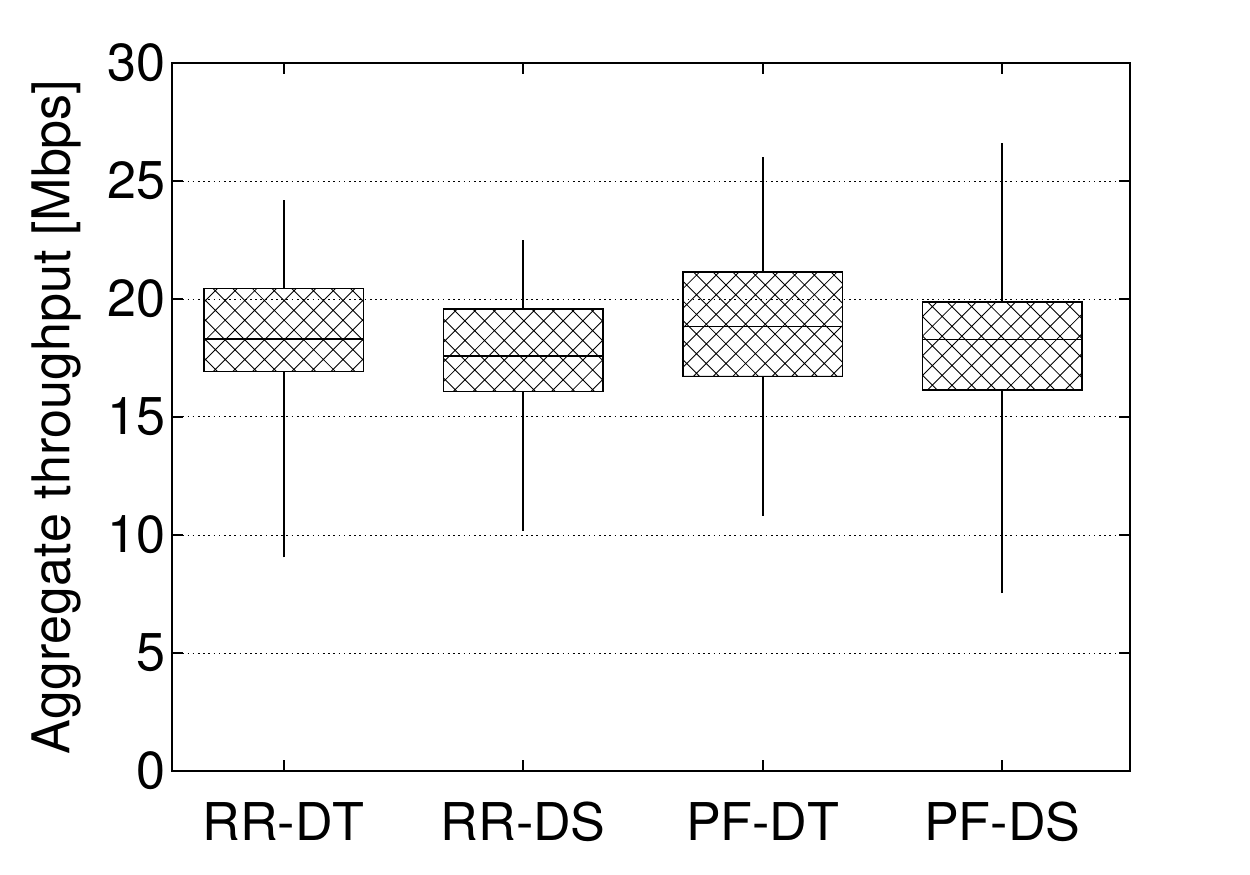} 
			\label{fig:QoSAggTput}
		}
		\subfigure[End to end delay.]
		{
			\includegraphics[scale=0.45, angle=0]{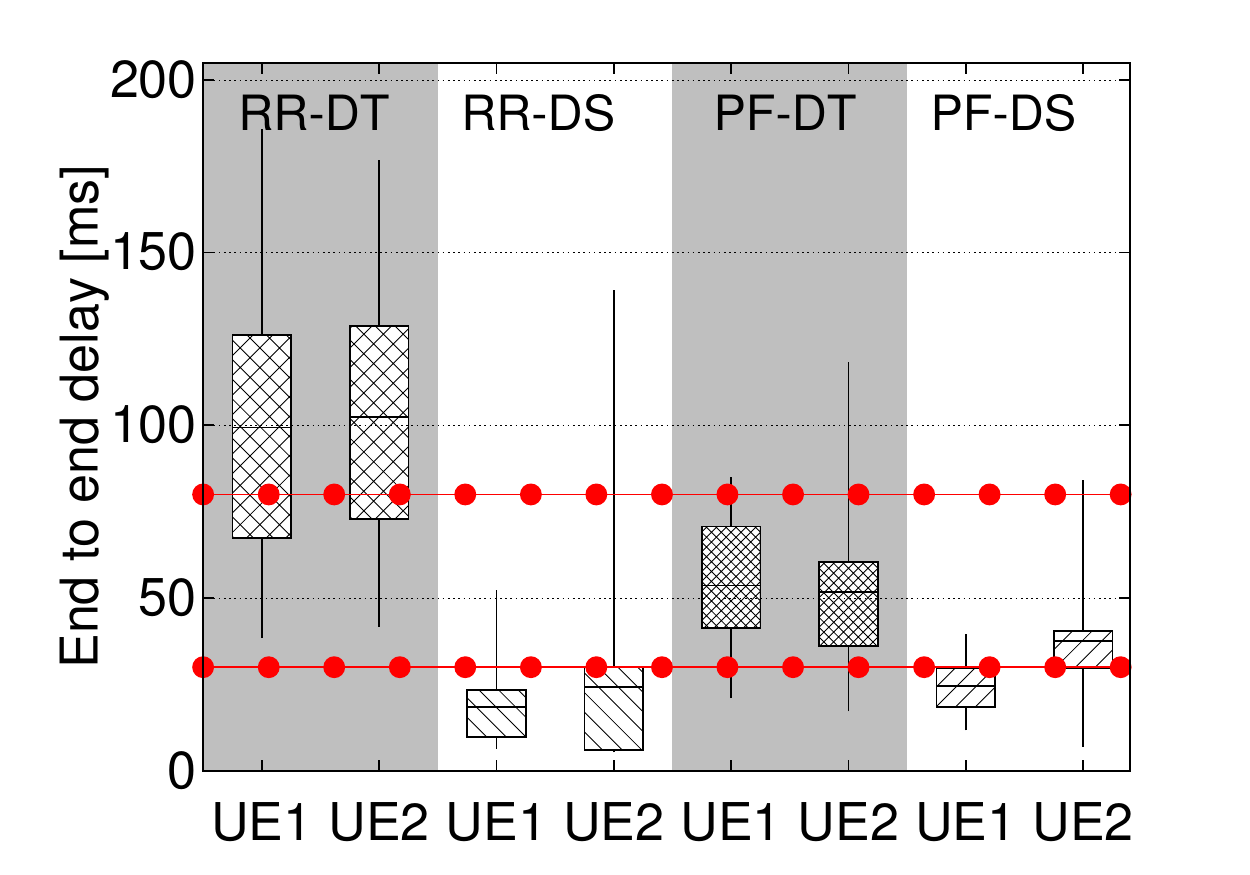} 
			\label{fig:QoSDelay}
		}
		\subfigure[Fairness.]
		{
			\includegraphics[scale=0.45, angle=0]{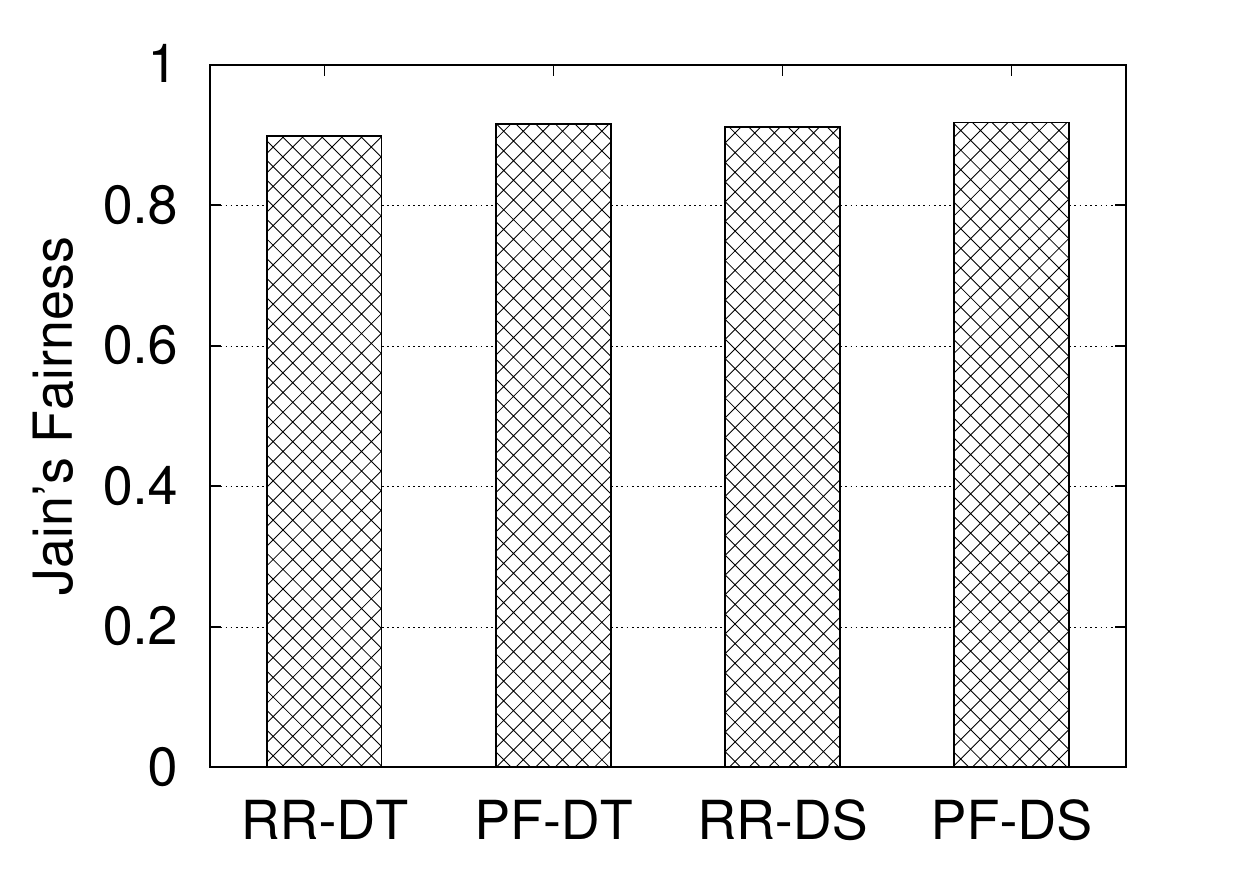} 
			\label{fig:QoSFair}
		}				
		\caption{Impact of QoS-awareness of DORE on system performance.}
		\label{fig:QoS}
\end{figure*}

Fig.~\ref{fig:QoSAggTput} shows the aggregate throughput of DORE with RR and PF but without QoS constraints (RR-DT and PF-DT in the figure) and with tight constraints (RR-DS and PF-DS). Both RR-DS and PF-DS achieve slightly lower throughput ($3\%$) w.r.t RR-DT and PF-DT because the QoS-awareness of DORE prevents opportunistic relay when delay constraints are violated. However, the $3\%$ throughput loss is a small price to pay to maintain the QoS requirements of the time-sensitive applications. 
Indeed, we observe in Fig.~\ref{fig:QoSDelay} that DORE can successfully cap the average delay below 30 ms and 80 ms. The effectiveness of DORE is especially seen when it reduces the packet delay of the voice traffic from $100$ ms to $23$ ms and $30$ ms. Since DORE delay control mechanism relies on UE feedbacks, it cannot avoid the delay caused by dramatic channel variations. As a result, the maximum delay under RR-DS and PF-DS can be higher than the delay thresholds.
The fairness performance of DORE with/without delay control mechanism is shown in Fig.~\ref{fig:QoSFair}. The results confirm that the delay control mechanism does not lead to unfairness among users.

\subsection{Quality of Experience (QoE) with DORE}
Good QoS does not necessarily correspond to good QoE. Thus, we design a video streaming scenario using VLC\footnote{http://www.videolan.org/vlc} to measure the QoE in terms of structural similarity. We use {\it AviSynth}\footnote{http://www.sourceforge.net/projects/avisynth2/files/AviSynth 2.5/} to measure structural similarity. Both PF and RR demonstrated similar trend hence we only show the result for PF, for brevity.  Again, we show in Fig.~\ref{fig:ssim} the structural similarity of the received video with $30$ ms delay constraint (i.e., PF-DS) and with an infinite one (i.e., PF-DT). We repeat the experiment for three different videos with 240p, 360p, and 480p resolutions. The results indicate that the QoS awareness of DORE results in up to $26\%$ structural similarity improvement. These values degrade with higher resolution videos because they are more sensitive to channel impairments. We also demonstrate a snapshot of the received video for 240p and 360p resolutions, in Fig.~\ref{fig:snapshot}. As expected, tight QoS constraints result in better image quality.

\begin{figure}[!h]
	\centering
		\subfigure[Structural similarity (SSIM) between streamed videos. ]
		{
			\includegraphics[scale=0.55, angle=0]{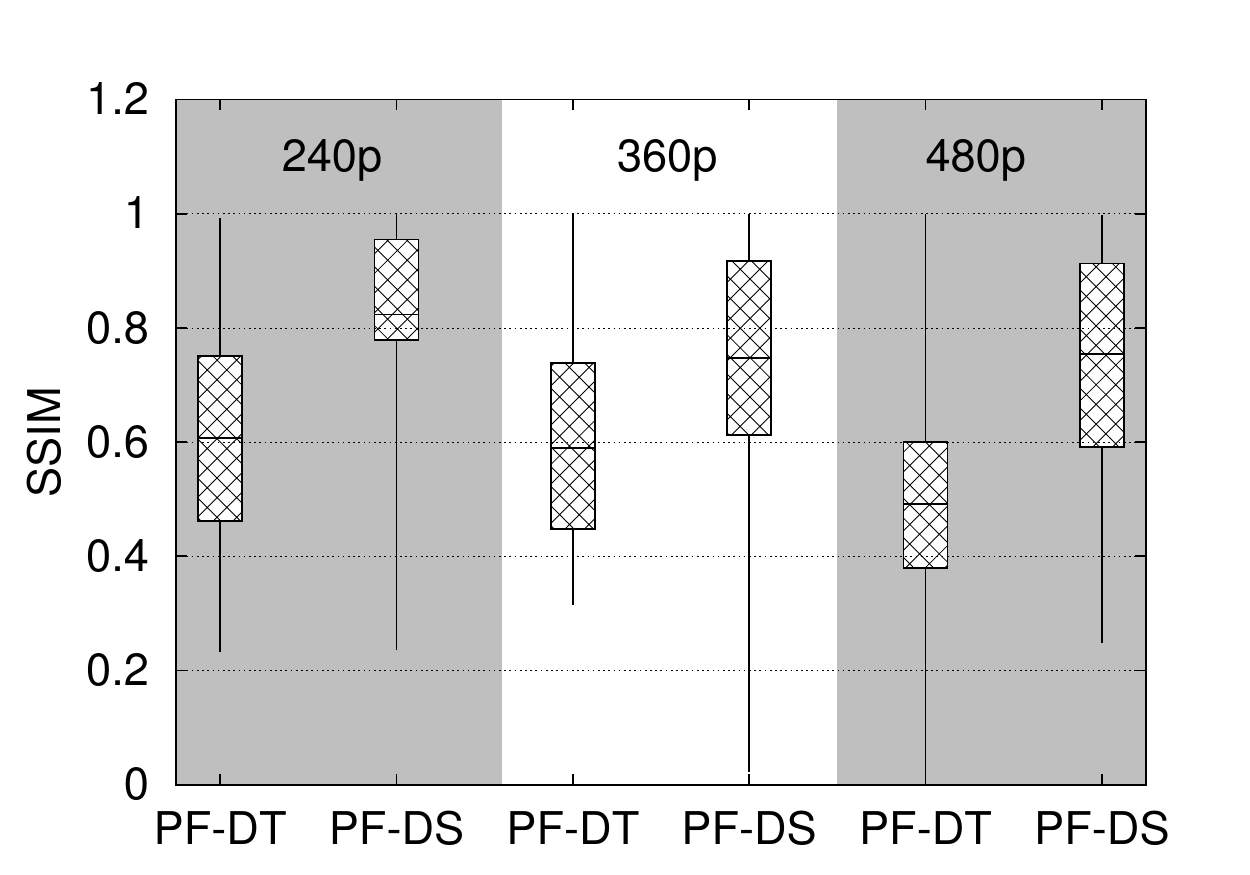} 
			\label{fig:ssim}
		}
		\subfigure[Snapshots of streamed video at the receiver.]
		{
			\includegraphics[scale=0.35, angle=0]{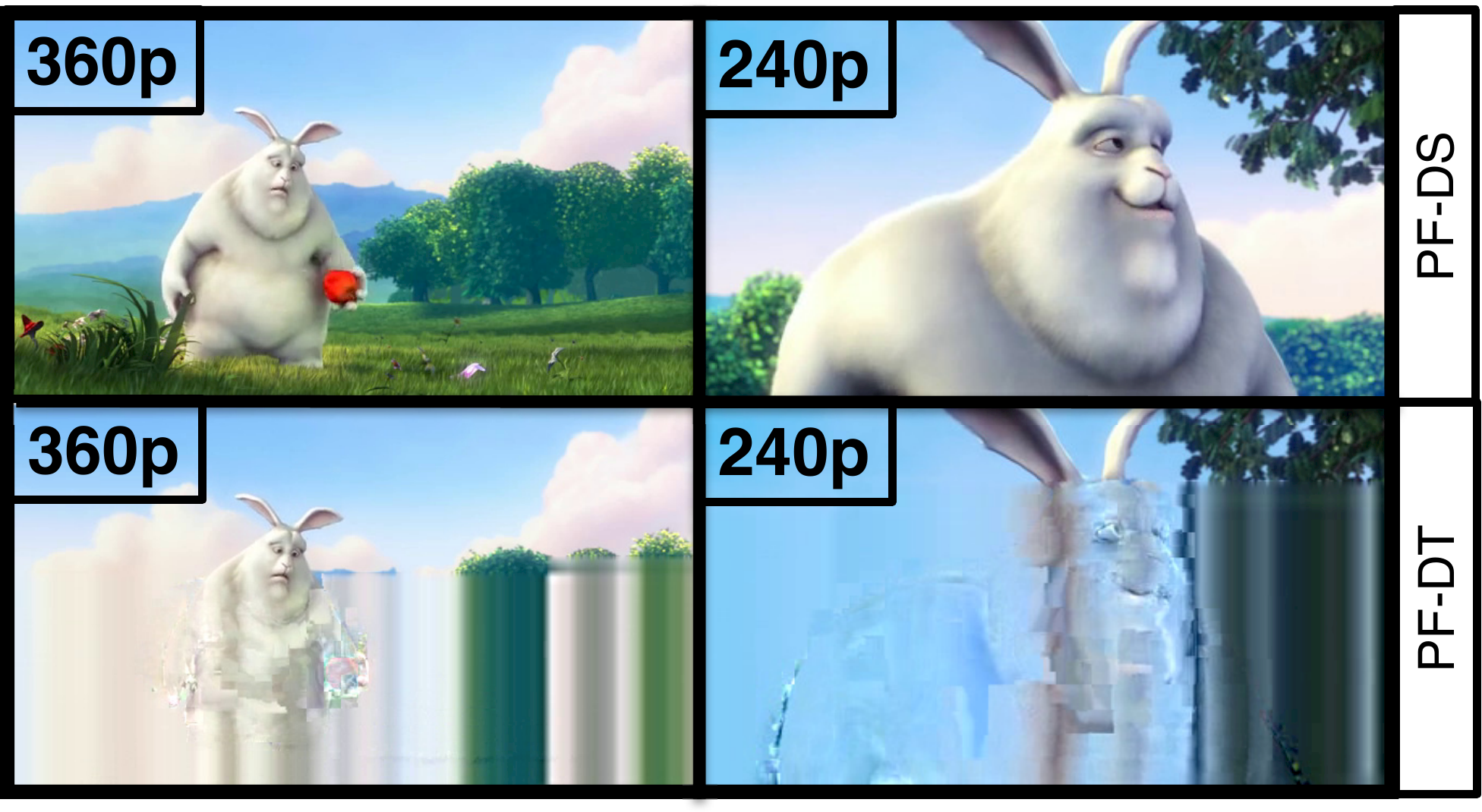} 
			\label{fig:snapshot}
		}		
		\caption{QoE performance of DORE.}
		\label{fig:video}
\end{figure}


\subsection{Opportunistic Relay within Large Relay Groups}
In the previous experiments, only two UEs were allowed to collaborate in DORE. Our observation in Fig.~\ref{fig:shadowTput} showed that the impact of opportunistic outband relay with only two users is limited. 
Since one-to-many communication is also present in 3GPP ProSe services, we can increase the size of the outband D2D group in order to achieve higher throughput. This experiment is designed to illustrate the impact of larger D2D UEs groups (see Fig.~\ref{fig:testbed}). Here, all UEs report their CQIs to the eNB, and the eNB performs the relay selection based on the reported channel qualities. The setup of this scenario is similar to \ref{ss:evalDore}, but all shadow UEs are allowed to join the D2D groups and can act as relays. Thus, the throughput of a shadow UE is measured at the eNB because they cannot decode messages produced by our eNB due to lack of compatibility between our experimental eNB and commercial smartphones. Note that UEs in the same group do not receive the same data, i.e., there is no multi-cast transmission in place in our scheme and experiments.
Fig.~\ref{fig:cluster} shows the aggregate system throughput. Our results confirm that by enlarging the outband D2D group from $2$ to $5$ UEs, the network throughput increases up to $71.8$\%. The result is critical to confirm the potentials of opportunistic D2D schemes. Indeed, we are the first to assess the opportunistic gain with multiple UEs relaying traffic among each other with a real implementation of an eNB scheduler and real-time CQI acquisition from multiple UEs. The reported results are obtained under PF scheduling. The achievable gains are even higher with RR, as shown in prior subsections. 

\begin{figure} [t!]
\centering
		\includegraphics[scale=0.60, angle=0]{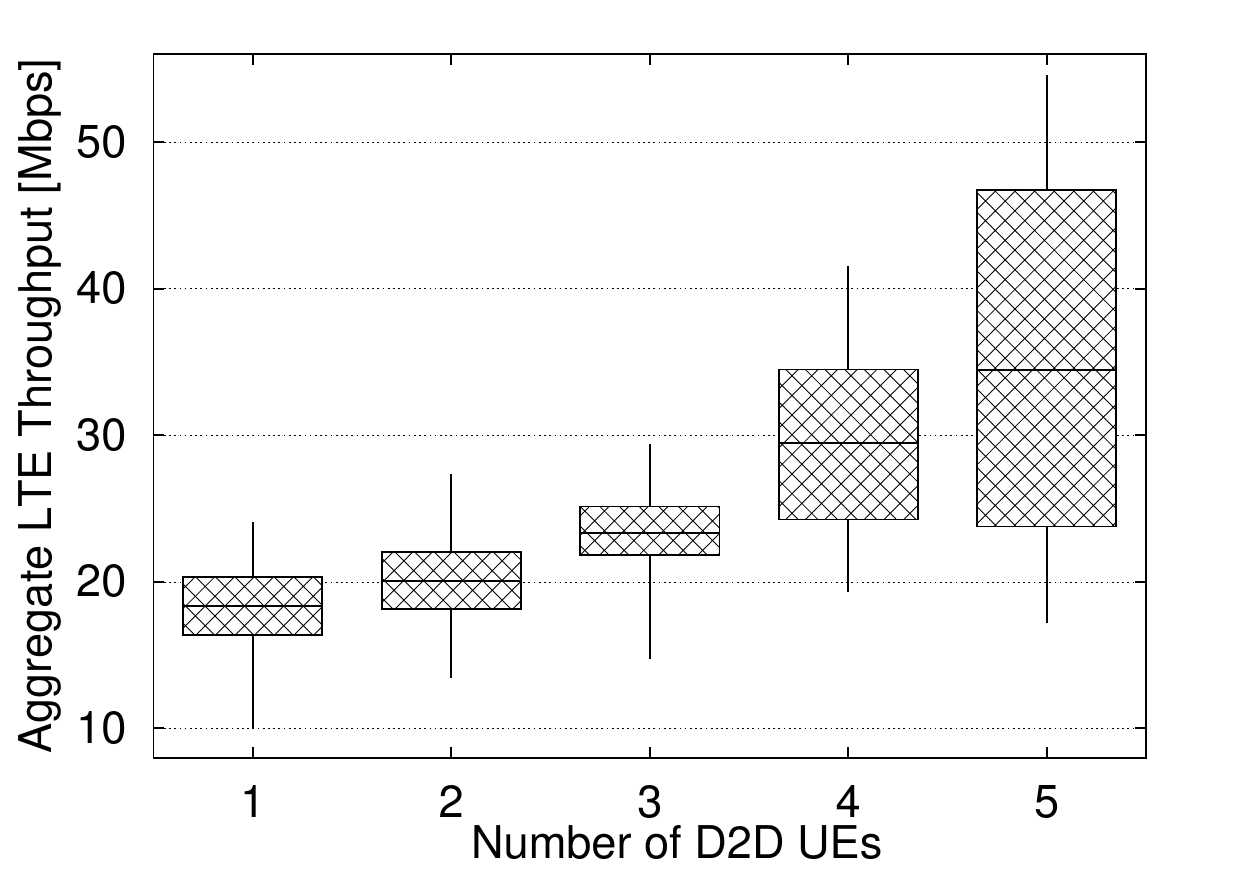} 
		\caption{Aggregate throughput versus the number of UEs in the same opportunistic outband D2D group. We observe that the opportunistic gain grows with the group size.}
		\label{fig:cluster}
		
\end{figure}

\vspace{-2mm}
\section{Discussion}
\vspace{-1mm}
\label{s:disc} 
This work provides in-depth intuitions to understand the practicality of integrating outband D2D communications in cellular networks. 
This section is dedicated to discuss the feasibility of such integration and to enlighten some key requirements for developing an experimental setup as well as for designing possible use-cases.

{\bf Feasibility.}
The SDR-based implementation of DORE is the proof-of-concept for the feasibility of outband D2D schemes with more complex and dynamic scenarios than non-opportunistic and QoS-unaware UE to UE communications. 

{\bf Implementation.}
There are several challenging issues to solve for SDR implementation of a D2D system. Here, we point out the most critical ones. The relay UE experiences high computational overhead due to LTE frame processing. Hence, we propose label switching at LTE PDCP layer instead of IP routing which is the current solution in 3GPP. As explained in Section~\ref{ss:ProSeamend}, the byproduct of this design choice is the elimination of relay-related security concerns. During the course of DORE implementation, we realized that D2D UEs switch role with high frequency (in order of milliseconds). Thus, we place DORE at the eNB instead of ProSe Function/Server to meet timing constraints and to avoid the additional overhead on the backhaul links.

{\bf Choice of the platform.}
To date, there are a few SDR platforms with {\it `simultaneous'} LTE and WiFi capability, namely, Open Air Interface, and LabVIEW. We choose LabVIEW for its modular and graphical programming structure that allows for quick real-time and FPGA code development without stepping into complex low-level programming languages. 
The choice of the NI PXI-based platform over USRP is due to the real-time capability of the PXI system that speeds up MAC layer algorithm prototyping and testing.

{\bf Capacity.}
DORE is key for boosting network capacity in one-to-many relay scenarios. This result is very promising, and it may suggest increasing the size of the relay groups as much as possible. However, in virtue of our observations on the extra load due to relay operations, it is plausible to suggest that each relay group should not include more than a handful of users, which is enough to enhance the network capacity by $70\%$. 

{\bf QoS.}
QoS provisioning is a concern in outband D2D due to the use of unlicensed spectrum. As a result, we designed DORE and the surrounding protocol with necessary feedback and handlers to enable QoS monitoring in our testbed. The experiments confirmed that DORE achieves the QoS requirements using a simple monitoring and feedback scheme.

{\bf Use-cases.}
Our experimental evaluation showed that (opportunistic) outband D2D schemes have low latency and ameliorate the throughput substantially. Hence, these schemes suit a large variety of applications including voice calls, video streaming, real-time gaming, and content sharing. 


\section{Related Work}
\label{s:related}

The literature on outband D2D both evaluates the potential performance gain using analysis/simulations and studies the feasibility of implementing outband D2D in today's cellular networks. We review the body of work in both groups. 
The authors of~\cite{kim2015TON} study the problem of efficient video delivery in D2D scenarios with quality-awareness. To this aim, the authors propose centralized and decentralized scheduling and streaming algorithms. They show via simulations that their algorithms can significantly outperform FlashLinQ~\cite{wu2013flashlinq} and Dynamic Adaptive Streaming over HTTP. In addition, the authors show that a well-designed distributed scheduling algorithm could perform very close to a centralized algorithm. An elaborated review of D2D-based video delivery schemes is provided in~\cite{ji2016JSAC}.

The authors of~\cite{asadi2014ComCom, golrezaei2012Globecom, bao2013Infocom} study the potential of the outband D2D relay. 
In our previous work~\cite{asadi2014ComCom}, we claim that 
the combination of opportunistic scheduling and outband D2D 
achieves 
$50$\% capacity gains in comparison to legacy cellular transmissions. 
Bao {\it el al.}~\cite{bao2013Infocom} propose the so-called Dataspotting approach that leverages outband D2D communications for content distribution in dense networks. Their proposal consists in using geo-location information of the content and its demand to offload part of the network load over the D2D links.
Golrezaei {\it et al.}~\cite{golrezaei2012Globecom} propose to use outband D2D and content caching techniques to improve video transmission in cellular networks 
by one or two orders of magnitude.

In~\cite{asadi2013WD, andreev2014ComMag, karvounas2014ComMag}, the authors investigate the necessary modifications to integrate LTE and WiFi to implement outband D2D. 
Andreev {\it et al.}~\cite{andreev2014ComMag} compare outband and inband D2D in terms of implementation complexity and their standardization progress. The authors conclude that the outband D2D has a higher implementation opportunity because inband D2D requires a significant change in the existing standard. 
In~\cite{karvounas2014ComMag} and~\cite{asadi2013WD}, the authors show that outband D2D can be implemented with minor modifications to the signaling procedure of LTE and group formation of WiFi Direct. 
In essence, these works point out that D2D is not a far-fetched concept anymore. Moreover, their studies reveal that outband D2D is a viable option for the first commercial implementation of D2D due to its simplicity in comparison to inband D2D. 

All the works above provide numerous analyses that aid towards the understanding of performance gains as well as the potential implications of D2D communications. Nevertheless, neither evaluation in testbed nor real-world implementation without {\it any simplifying assumptions} has been investigated for outband D2D systems, in contrast to our work. 
This article is an extended version of~\cite{asadi2016INFOCOM} in which we present a more thorough problem motivation based on real-time LTE-A channel measurements and in which, in addition, our experiments are extended to show fundamental KPIs such as fairness achieved by the users and CPU load caused by the dynamic control of D2D relay.

\section{Conclusions}
\label{s:conclusions} 


In this paper, we prototyped the first SDR platform for outband D2D communications. We leveraged Xilinx FPGAs and the NI Real-Time OS to develop realistic experiments with LTE-like millisecond CQI reporting, scheduling, and high-speed LTE-WiFi interaction. Our experimental evaluation using several QoS and QoE metrics confirmed the feasibility and potentials of opportunistic outband D2D communications. In particular, we designed DORE which is a 3GPP ProSe-compliant and QoS-aware opportunistic outband D2D framework. The results revealed that experimental performance figures are lower than the reported values in the prior analytical studies, although still notable (up to $20$\% with just two users). Nevertheless, high throughput gains are achievable if the number of participating UEs in opportunistic outband D2D increases (up to $71$\% with five users). Moreover, our experiments corroborated the efficiency and robustness of DORE for both delay tolerant and delay sensitive applications by maintaining the end-to-end delay below $50$ ms.

This work is the first step toward experimental examination of outband D2D communications. As such, there are many other research avenues that we plan to explore in the future.  As a key example, although our analytical/simulation results have proven the energy efficiency of outband D2D-relay, there is still no experimental study on the topic~\cite{asadi2014ComCom}. However, due to the use of PXI platforms, whose power consumption is much higher than smartphones, we could not conduct a meaningful energy measurement campaign. Nonetheless, we are currently working on migrating our testbed to USRP FlexRIO platforms so that we can study the impact of our proposal on energy consumption. Moreover, in this work, we have not evaluated the D2D discovery phase due to hardware limitations. However, it would be interesting to observe the impact of interference from other devices on the discovery phase, especially using autonomous discovery in very dense networks~\cite{asadi2015Commag, xenakis2016TWC}. Finally, the study of the co-existence of all D2D modes (outband and inband) is another interesting future research direction. This is in particular interesting in the context of 5G mmWave communication as mmWave properties such as directionality and short range can indeed enable the use of a virtually unbounded number of concurrent D2D links~\cite{sim20165g, qiao2015enabling, sim2017opp}.

\section{Acknowledgements}
This work has been partially supported by the Madrid Regional Government through the TIGRE5-CM program (S2013/ICE-2919), the LOEWE initiative (Hessen, Germany) within the NICER project, and by the German Research Foundation (DFG) in the Collaborative Research Center 1053 MAKI. This research was also supported by the Spanish Ministry of Economy and Competitiveness under grants TEC2014-55713-R and RYC-2014-01335 and by the European Commission in the framework of the H2020-ICT-2014-2 project Flex5Gware (Grant agreement no. 671563). The authors would like to thank National Instruments for the support in LabVIEW FPGA code development.


\bibliographystyle{IEEEtran}


\bibliography{biblio}

\vspace{-20mm}
%

\end{document}